\documentclass[times, twosid]{StyleBioRxiv}
\usepackage{blindtext}
\usepackage{flafter}
\usepackage{booktabs}
\usepackage[switch]{lineno}

\leadauthor{Dworkin} 

\begin{document}
\setlength\parindent{18pt}

\title{The extent and drivers of gender imbalance in neuroscience reference lists}
\shorttitle{Neuroscience citation bias}

\author[1]{Jordan D. Dworkin}
\author[1]{Kristin A. Linn}
\author[2]{Erin G. Teich}
\author[3]{Perry Zurn}
\author[1]{\\Russell T. Shinohara}
\author[2,4,5,6,7,8,\Letter]{Danielle S. Bassett}

\affil[1]{Penn Statistics in Imaging and Visualization Center, Department of Biostatistics, Epidemiology, and Informatics, Perelman School of Medicine, University of Pennsylvania, Philadelphia, PA, USA}
\affil[2]{Department of Bioengineering, School of Engineering and Applied Science, University of Pennsylvania, Philadelphia, PA, USA}
\affil[3]{Department of Philosophy and Religion, American University, Washington DC, USA}
\affil[4]{Department of Physics \& Astronomy, College of Arts and Sciences, University of Pennsylvania, Philadelphia, PA, USA}
\affil[5]{Department of Electrical \& Systems Engineering, School of Engineering and Applied Science, University of Pennsylvania, Philadelphia, PA, USA}
\affil[6]{Department of Neurology, Perelman School of Medicine, University of Pennsylvania, Philadelphia, PA, USA}
\affil[7]{Department of Psychiatry, Perelman School of Medicine, University of Pennsylvania, Philadelphia, PA, USA}
\affil[8]{Santa Fe Institute, Santa Fe, NM, USA}

\maketitle

\begin{abstract}
\noindent Like many scientific disciplines, neuroscience has increasingly attempted to confront pervasive gender imbalances within the field. While much of the conversation has centered around publishing and conference participation, recent research in other fields has called attention to the prevalence of gender bias in citation practices. Because of the downstream effects that citations can have on visibility and career advancement, understanding and eliminating gender bias in citation practices is vital for addressing inequity in a scientific community. In this study, we sought to determine whether there is evidence of gender bias in the citation practices of neuroscientists. Using data from five top neuroscience journals, we find that reference lists tend to include more papers with men as first and last author than would be expected if gender were not a factor in referencing. Importantly, we show that this overcitation of men and undercitation of women is driven largely by the citation practices of men, and is increasing over time as the field becomes more diverse. We develop a co-authorship network to assess homophily in researchers’ social networks, and we find that men tend to overcite men even when their social networks are representative. We discuss possible mechanisms and consider how individual researchers might address these findings in their own practices.
\end{abstract}


\begin{corrauthor}
dsb\at seas.upenn.edu
\end{corrauthor}


\section*{Introduction}
In recent years, science has been pushed to grapple with the social and structural systems that produce vast gender imbalances in academic participation. Research has found large and persistent gaps in the proportion of women across scientific fields and has estimated that many fields will not reach gender equity for decades at their current trajectories \citep{holman_gender_2018}. For women currently or formerly in academia, gender imbalances have persisted across various measures of academic inclusion and success. Prior work has found that such inequalities are present in compensation \citep{moss-racusin_science_2012}, grant funding \citep{bornmann_gender_2007,jagsi_sex_2009,van_der_lee_gender_2015}, credit for collaborative work \citep{sarsons_recognition_2017}, teaching evaluations \citep{macnell_whats_2015,mengel_gender_2019,boring_gender_2017}, hiring and promotions \citep{nielsen_limits_2016,van_den_brink_scouting_2011,de_paola_gender_2015}, productivity and authorship \citep{west_role_2013,wilhelm_new_2018,lariviere_bibliometrics:_2013,huang_historical_2020}, and citations \cite{ferber_gender_2011,maliniak_gender_2013,caplar_quantitative_2017}. Importantly, while this study focuses on gender, similar biases have been demonstrated in domains like race, socioeconomic status, and university prestige \citep{fang_racial_2000,petersen_reputation_2014,way_productivity_2019}.

While many aspects of gender bias have yet to be studied within neuroscience specifically, issues of gender and diversity have commanded increasing attention over the past several years. Groups like BiasWatchNeuro (biaswatchneuro.com), Women in Neuroscience (winrepo.org), and Anne’s List (anneslist.net) have been created to track and promote the inclusion of women in conferences and symposia. Furthermore, major neuroscience societies have publicly discussed ways to improve representation \citep{joels_tale_2014}, and journals have sought to balance the composition of editors and reviewers \citep{noauthor_promoting_2018}. On the heels of these efforts, a recent study showed that authorship and public speaking have indeed become more balanced in the last decade \citep{schrouff_gender_2019}. 

However, measures of authorship and conference participation reflect only one aspect of success in a field, and the presence of differential engagement with scholarship could lead to prolonged inequities in other areas. Recent studies of such differential engagement have found not only that people from marginalized groups are broadly undercited in fields such as communications \citep{chakravartty_communicationsowhite_2018} and philosophy \citep{thiem_just_2018}, but also that women-led research in particular tends to receive fewer citations than comparable papers led by men in the fields of astronomy \citep{caplar_quantitative_2017}, international relations \citep{maliniak_gender_2013}, and political science \citep{dion_gendered_2018}. Theoretical work has proposed a “Matilda effect” in which the contributions of men are seen as more central within a field and are therefore sought out more often and evaluated more highly \citep{rossiter_matthew_1993}. In visual art and literary texts, the “Bechdel test” has revealed the prevalence of cases in which women’s contributions are not valued independently of men's \citep{selisker_bechdel_2015,garcia_asymmetries2014}. The presence of such an effect in scientific authorship would likely produce reputational and citational inequity. In this case, women-led work could remain under-discussed and perceived as marginal to men-led work.

Because of the potential for harmful downstream effects of inequitable engagement with women and men's work, the study of citation behavior is a critical endeavor for understanding and addressing a field’s biases. Additionally, achieving gender equity within citation lists is a goal that can be pursued by all researchers during their paper-writing process (unlike, for example, achieving gender equity within keynote speaker roles). Thus, in this study we seek to determine the existence and potential drivers of gender bias in neuroscience citations. Previous work in citation gaps has often focused on the relationship between authors’ gender and their citation counts \citep{caplar_quantitative_2017, maliniak_gender_2013}, finding that work by women tends to receive fewer citations than similar work by men. Yet this formulation only measures the passive consequences of gendered citation behavior, rather than directly measuring the behavior itself. Instead, building on recent studies conducted in international relations and political science \citep{mitchell_gendered_2013,dion_gendered_2018}, we investigate the relationship between authors’ gender and the gender make-up of their reference lists. Using this framework, we are able to quantify properties associated with authors serving as both objects and agents of undercitation.

For this study, we examine the authors and reference lists of articles published in five top neuroscience journals since 1995. Within this pool of articles, we are able to obtain probabilistic estimates of authors’ gender identity, find connections between citing and cited papers, locate and remove instances of self-citation, and study the links between authors’ genders and their role as objects/agents of undercitation. Specifically, we test the following hypotheses: (1) The overall citation rate of women-led papers (defined here as those with women as first- and/or last- author) will be lower than expected given the papers’ relevant characteristics; (2) The undercitation of women-led papers will occur to a greater extent within men-led reference lists; (3) Undercitation of women-led papers will be decreasing over time, but at a slower rate within men-led reference lists; (4) Differences in undercitation between men-led and women-led reference lists will be partly explained by the structure of authors’ social networks. Significance will be assessed for these hypotheses using a null model that preserves the structure of the citation graph, and all \textit{p}-values will be corrected for multiple comparisons.

\section*{Results}
\subsection*{Data description}
Using Web of Science, we extracted data on research articles, reviews, and proceedings published in five top neuroscience journals since 1995. We selected the journals \textit{Nature Neuroscience}, \textit{Neuron}, \textit{Brain}, \textit{Journal of Neuroscience}, and \textit{NeuroImage}, as they were reported by the Web of Science to have the highest Eigenfactor scores \citep{bergstrom_eigenfactortm_2008} among journals in the neuroscience category. In all, 61,416 articles were included in the final dataset of citing/cited papers. Full author names were provided by Web of Science for all articles published after 2006. For all articles published in 2006 or earlier, full names were drawn, when available, from Crossref or the journals’ websites. To minimize missing data, we developed an algorithm to match authors for whom only first/middle initials were available to other authors in the dataset with the same initials and last name (see Methods). 

Gender was assigned to first names using the `gender' package in R \citep{blevins_jane_2015} with the Social Security Administration (SSA) baby name dataset. For names that were not included in the SSA dataset, gender was assigned using Gender API (gender-api.com), a paid service that supports roughly 800,000 unique first names across 177 countries. We assigned `man'(`woman') to each author if their name had a probability greater than or equal to 0.70 of belonging to someone labeled as `man'(`woman') according to a given source \citep{dion_gendered_2018}. In the SSA dataset, man/woman labels correspond to the sex assigned to children at birth; in the Gender API dataset, man/woman labels correspond to a combination of sex assigned to children at birth and genders detected in social media profiles. In a random sample of 200 authors, the accuracy of these automated assignments was 0.96 (see Supplementary Information and Tables \ref{tab:tabvalaut}-\ref{tab:tabvalpap} for further details). Gender could be assigned to both the first and last author of 88\% of the papers in the dataset. Of the 12\% of papers with missing data, 7\% were missing because either the first- or last-author’s name had uncertain gender, and 5\% were missing because either the first- or last-author’s name was not available. We performed the following analyses using the articles for which gender could be assigned with high probability to both authors (n = 54,225), but sensitivity analyses conducted on the full data can be found in the Supplementary Information (see Table \ref{tab:tabmiss}).

In gender theory, sex often refers to physical attributes, as determined anatomically and physiologically, while gender often refers to a self-identity, as expressed behaviorally and in sociocultural context \citep{fausto-sterling_sexing_2000}. In our analysis, the term “gender” does not refer directly to the sex of the author, as assigned at birth or chosen later, nor does it refer directly to the gender of the author, as socially assigned or self-chosen. The term “gender,” in our analysis, is a function of the probability of assigned gendered names. By “woman,” we mean an author whose name has a probability greater than or equal to 0.70 of being given to a child assigned female at birth or belonging to someone identifying as a woman on social media; likewise, by “man,” we mean an author whose name has a probability greater than or equal to 0.70 of being given to a child assigned male at birth or belonging to someone identifying as a man on social media. The author’s actual sex or gender is not identified. 

Given the limitations of both probabilistic analyses and of birth assignments, the authors may in fact have a sex or gender different from the one we have assigned and/or be intersex, transgender, or nonbinary \citep{feder_making_2014,stryker_transgender_2008}. In some cases, citers will know the sex and/or gender of the authors they cite. In many cases, they will not know but rather infer, often via a name, the gender of the authors they cite. Instances of both known and inferred gender have the potential to incite either explicit or implicit bias in citing authors \citep[i.e., where explicit bias involves conscious cognitive processing, implicit bias is automatic cognitive processing that presupposes social prejudices and stereotypes;][]{greenwald_measuring_1998,brewer_psychology_1999,brownstein_implicit_2019}. Our probabilistic analysis by gendered name therefore functions to nontrivially capture bias arising due to both known and inferred gender in citation practices.

\begin{figure*}[ht]
\centering
\includegraphics[width=.9\linewidth]{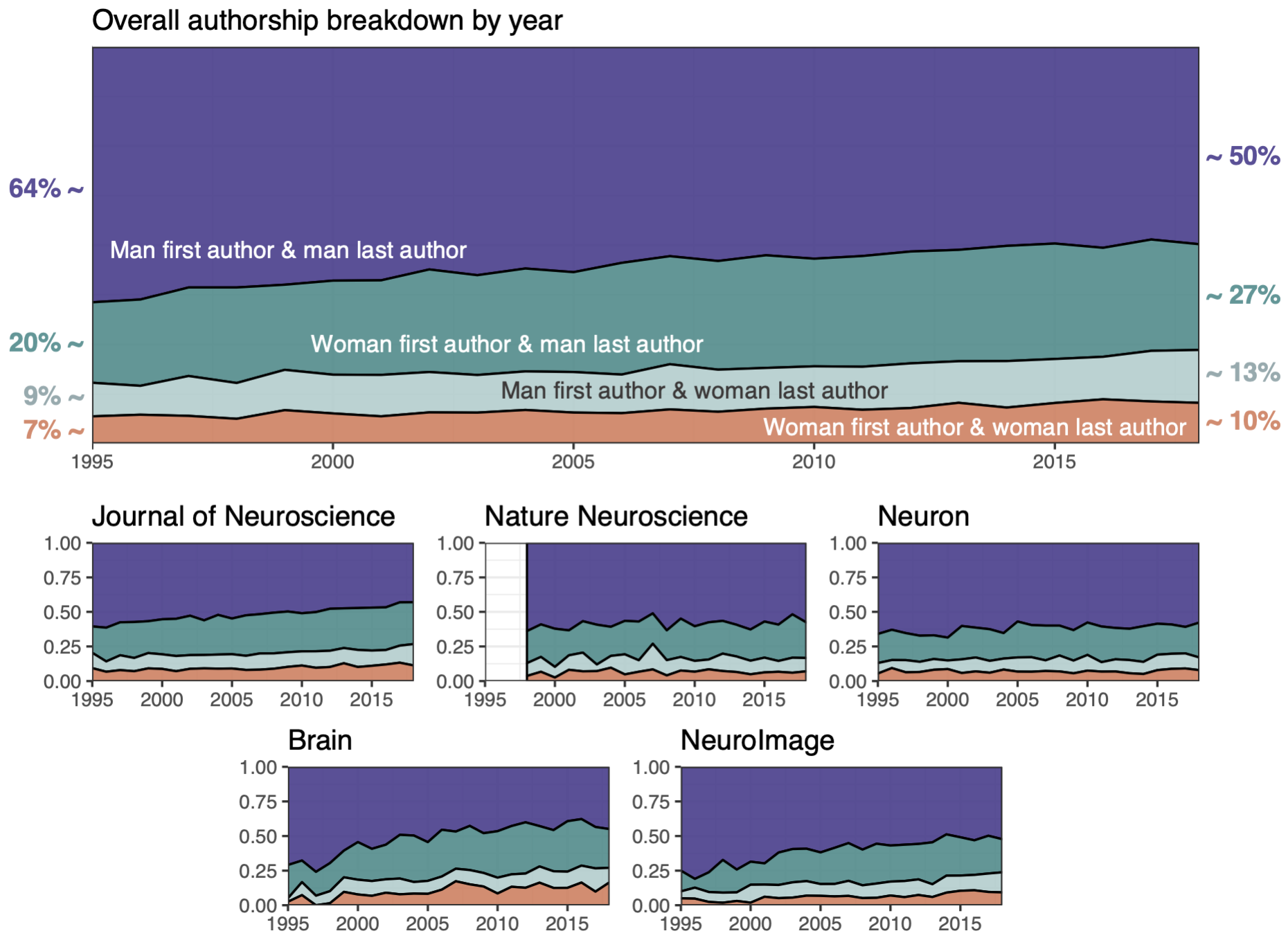}
\caption{\textbf{Trends in author gender within top neuroscience journals between 1995 and 2018.} Top panel shows the overall trends across the five journals studied. Bottom panels show the trends within each journal. From top to bottom, panels show the proportion of articles with men as first and last author (purple), women as first author and men as last author (green), men as first author and women as last author (gray), and women as both first and last author (orange). Note: \textit{Nature Neuroscience} was not established until 1998.}
\label{fig:Figure1}
\end{figure*}

\subsection*{Trends in authorship}
Across the articles in the sample, the proportion of articles with a woman as first or last author significantly increased between 1995 and 2018, at a rate of roughly 0.60\% per year (95\% CI = [0.53, 0.67]). This trend varied across journals, with the \textit{Journal of Neuroscience} (0.67; 95\% CI = [0.57, 0.77]), \textit{NeuroImage} (0.89; 95\% CI = [0.72, 1.06]), and \textit{Brain} (1.16; 95\% CI = [0.92, 1.41]) all showing increases between 0.65\% and 1.2\% per year. \textit{Neuron} showed a modest increase of 0.29\% per year (95\% CI = [0.12, 0.45]), and \textit{Nature Neuroscience} did not show a clear increasing trend (0.19; 95\% CI = [-0.09, 0.46]). Across these five journals, the overall proportion of articles that were either first- or last-authored by women increased from 36\% in 1995 to 50\% in 2018 (Figure \ref{fig:Figure1}).

\subsection*{Citation imbalance relative to overall authorship proportions}
To quantify citation behavior within neuroscience articles, we specifically examined the reference lists of papers published between 2009 and 2018 (n = 31,418). Thus, while all papers in the dataset were potential \textit{cited} papers, references to \textit{citing} papers refer only to those published since 2009. For each citing paper, we took the subset of its citations that had been published in one of the above five journals since 1995 and determined the gender of the cited first and last authors. We removed self-citations (defined as cited papers for which either the first or last author of the citing paper was first-/last-author; see Methods for further detail) from consideration for all analyses presented in the main text, but see the Supplementary Information for detailed analyses of the role of self-citations. We then calculated the number of cited papers that fell into each of the four first author \& last author categories: man \& man (MM), woman \& man (WM), man \& woman (MW), and woman \& woman (WW). Single-author papers by men and women were included in the MM and WW categories, respectively.

\begin{figure*}[ht]
\centering
\includegraphics[width=1\linewidth]{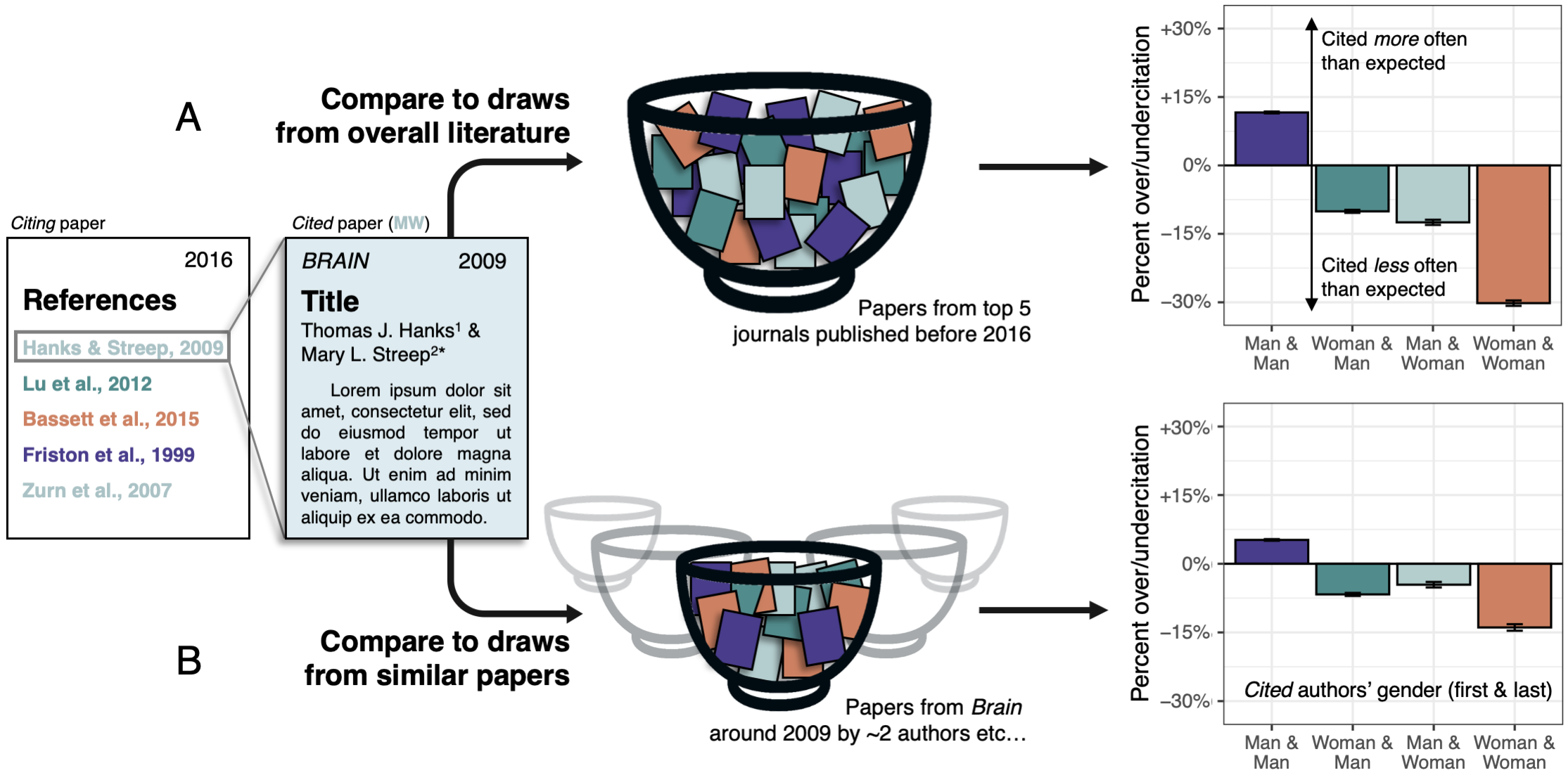}
\caption{\textbf{Construction and visualization of over/undercitation of papers based on author gender.} \textit{(A)} Illustration of the random draws model, in which gender proportions in reference lists are compared to the overall gender proportions of the existing literature. Right panel shows the over/undercitation of different author gender groups compared to their expected proportions under the random draws model. \textit{(B)} Illustration of the relevant characteristics model, in which gender proportions in reference lists are compared to gender proportions of articles that are similar to those that were cited across various domains. Right panel shows the over/undercitation of different author gender groups compared to their expected proportions under the relevant characteristics model.}
\label{fig:Figure2}
\end{figure*}

As a simple first step, we compared the observed number of citations within each category to the number that would be expected if references were drawn randomly from the pool of papers (Figure \ref{fig:Figure2}A). To obtain the number that would be expected under this assumption of random draws, we calculated the gender proportions among all papers published prior to the citing paper – thus representing the proportion among the pool of papers that the authors could have cited – and multiplied them by the number of papers cited. The following section expands this naive measure to account for potential relationships between author gender and other relevant characteristics of cited papers.

Of the 303,886 citations given between 2009 and 2018, MM papers received 61.7\%, compared to 23.6\% for WM papers, 9.0\% for MW papers, and 5.8\% for WW papers. The expected proportions based on the pool of citable papers were 55.3\% for MM, 26.2\% for WM, 10.2\% for MW, and 8.3\% for WW. We defined a measure of over/undercitation as the (observed \% - expected \%)/expected \% (see Methods for further details). This measure thus represents the percent over/undercitation relative to the expected proportion. By this measure, MM papers were cited 11.6\% more than expected (95\% CI = [11.2, 12.0]), WM papers were cited 10.1\% less than expected (95\% CI = [-10.7, -9.5]), MW papers were cited 12.5\% less than expected (95\% CI = [-13.6, -11.4]), and WW papers were cited 30.2\% less than expected (95\% CI = [-31.3, -29.0]). This set of percentages correspond to MM papers being cited roughly 19,500 more times than expected, WM papers being cited roughly 8,000 fewer times than expected, MW papers being cited roughly 3,900 fewer times than expected, and WW papers being cited roughly 7,600 fewer times than expected.

\subsection*{Citation imbalance after accounting for papers’ relevant characteristics}
The comparison of citations to overall authorship proportions does not take into account other important properties of published papers that may make them more or less likely to be cited by later scholarship. The potential relationship between author gender and papers’ other relevant characteristics makes it difficult to isolate the effects of gender on the rates at which work is cited. To address this issue, we developed a method for calculating the probabilities that a given citation would be for a {MM, WM, MW, WW} paper conditional on various salient characteristics of the cited paper. The characteristics of a paper that we selected as being potentially relevant for citation rates were 1) the year of publication, 2) the journal in which it was published, 3) the number of authors, 4) whether the paper was a review article, 5) the seniority of the paper's first and last authors. We then sought to compare the true citation rates to the rates that would be expected if only these non-gender characteristics were relevant.

We obtained the estimated gender probabilities by specifying a generalized additive model (GAM) on the multinomial outcome of paper authorship in the four specified categories of first and last author gender. Within the GAM framework, papers’ membership among these four categories was regressed on the characteristics described above (i.e., publication date, journal, author count, binary review article status, and first-/last-author seniority; see Methods for further details). Since seniority is a somewhat ambiguous concept, and is not defined in the available data, we defined authors’ seniority as the number of papers on which they had been a first or last author in the time span of the study (1995-2018). Thus, the estimated membership obtained for a specific article – given by the model as a set of four probabilities that sum to 1 – approximately represents the proportion of similar papers (i.e., same journal, published around the same time, etc.) that fall within each of the four gender categories.

To then estimate the gendered citation behavior of recent articles, accounting for the other relevant characteristics of cited papers, we compared the authorship gender category of each cited paper to its probabilities of belonging to each of the four categories. As opposed to the previous section, in which gender probabilities model citation as a random draw from the existing literature, the current probabilities can be viewed as the expected gender proportions across random draws from a narrow pool of papers highly similar to the cited paper (see Figure \ref{fig:Figure2}B). Interestingly, these probabilities can also be framed as the expected proportions if the genders of a cited paper's authors were randomly swapped across highly similar papers. Though we find the first framing slightly more intuitive, this second framing makes clear that the presented results do not depend on breaking the structure of the citation graph. In fact, the graph-preserving null model used to assess significance is based on this second framing (see Methods for further detail).

Summing up the number of cited papers from each category again gives us the observed citation rates, and summing up the authorship gender probabilities across the cited papers gives us the new expected citation rates. As reported above, MM papers received 61.7\% of citations, compared to 23.6\% for WM papers, 9.0\% for MW papers, and 5.8\% for WW papers. Based on the relevant properties of cited papers, the expected proportions were 58.6\% for MM, 25.3\% for WM, 9.4\% for MW, and 6.7\% for WW. Thus, after accounting for salient non-gender characteristics, MM papers were still cited 5.2\% more than expected (95\% CI = [4.8, 5.5], $p<0.001$), WM papers were cited 6.7\% less than expected (95\% CI = [-7.3, -6.0], $p=0.008$), MW papers were cited 4.6\% less than expected (95\% CI = [-5.7, -3.3], $p=0.86$), and WW papers were cited 13.9\% less than expected (95\% CI = [-15.2, -12.5], $p=0.003$). Of 303,886 total citations, these proportions correspond to citations being given to MM papers roughly 9,300 more times than expected, WM papers roughly 5,100 fewer times than expected, MW papers roughly 1,300 fewer times than expected, and WW papers roughly 2,800 fewer times than expected. The observed overcitation of MM papers and undercitation of WM and WW papers provides support for the hypothesis that the citation rate of women-led papers is lower than expected given relevant characteristics (Hypothesis 1).

\begin{figure*}[ht]
\centering
\includegraphics[width=.85\linewidth]{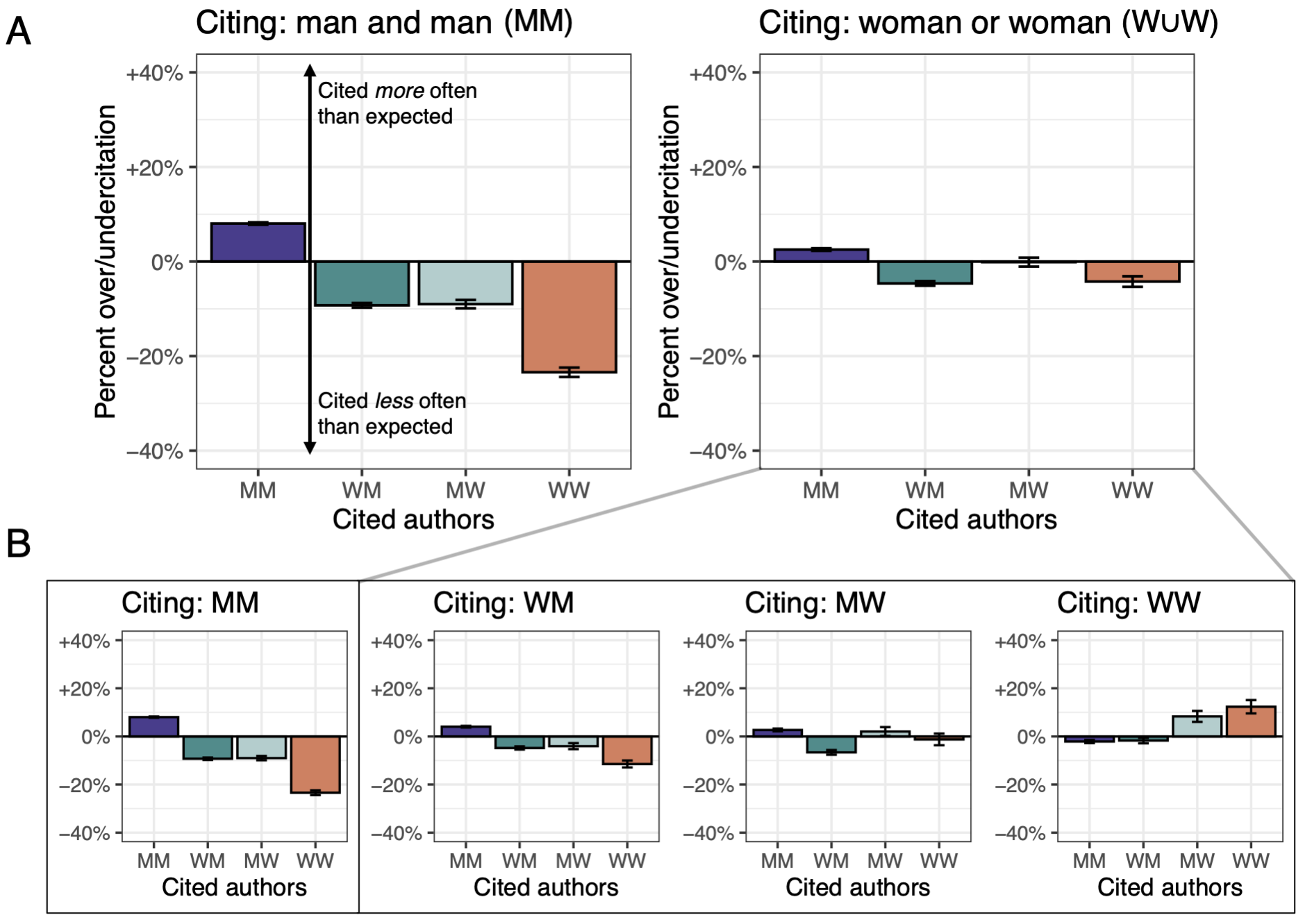}
\caption{\textbf{Relationship between author gender and gendered citation practices.} \textit{(A)} Degree of over/undercitation of different author genders within MM reference lists (left) and within W$\cup$W reference lists (right). Shows that papers with men as both first and last author overcite men to a greater extent than papers with women as either first or last author. \textit{(B)} Full breakdown of gendered citation behavior within MM, WM, MW, and WW reference lists.}
\label{fig:Figure3}
\end{figure*}

\subsection*{The effect of authors’ gender on citation behavior}
By focusing the present analyses on the gender make-up of reference lists, as opposed to the number of citations that articles receive, we are able to investigate the gender of the citing authors in addition to that of the cited authors. Thus, in this section we compare the gender make-up of references within papers that had men as both first and last author (referred to, as usual, as MM) to those within papers that had women as either first or last author (henceforth referred to as W$\cup$W, comprising WM, MW, and WW papers). Of the 31,418 articles published in one of the five journals between 2009 and 2018, roughly 51\% were MM and 49\% were W$\cup$W.

After separating citing articles by author gender, we find that the imbalance within reference lists shown previously is driven largely by the citation practices of MM teams. Specifically, within MM reference lists, other MM papers were cited 8.0\% more than expected (95\% CI = [7.6, 8.5], $p<0.001$), WM papers were cited 9.3\% less than expected (95\% CI = [-10.2, -8.3], $p<0.001$), MW papers were cited 9.0\% less than expected (95\% CI = [-10.6, -7.2], $p=0.10$), and WW papers were cited 23.4\% less than expected (95\% CI = [-25.5, -21.7], $p<0.001$; Figure \ref{fig:Figure3}A, left). Within W$\cup$W reference lists, MM papers were cited only 2.5\% more than expected (95\% CI = [2.0, 3.1], $p=0.07$), WM papers were cited 4.6\% less than expected (95\% CI = [-5.7, -3.7], $p=0.12$), MW papers were cited 0.1\% less than expected (95\% CI = [-2.0, 1.8], $p>0.99$), and WW papers were cited 4.2\% less than expected (95\% CI = [-6.5, -1.9], $p>0.99$; Figure \ref{fig:Figure3}A, right). The observed differences between MM and W$\cup$W reference lists were all significant ($p<0.0001$), providing support for the hypothesis that the undercitation of women-led papers occurs to a greater extent within men-led reference lists (Hypothesis 2).

Within the W$\cup$W group, the citation proportions of the WM, MW, and WW subgroups suggest a more fine-grained link between the increased citation of women-led work and the increased leadership role of women on the citing team (Figure \ref{fig:Figure3}B). Specifically, WM teams still slightly undercite WW papers relative to expectation, but do so at roughly half the rate of MM teams. MW reference lists contain roughly the expected citation proportions across gender groups, and WW reference lists contain slightly more MW and WW papers than expected (overciting WW papers at roughly half the rate that MM teams undercite WW papers). This moderate overcitation of women-led work within women-led reference lists points to a potential role of social networks in forming authors' mental representations of the available citable work. This possibility is explored in detail in a later section.

Two additional considerations merit brief discussion. First, research subfields are not accounted for when calculating expected gender proportions. To determine the extent to which this feature impacts the presented results, we conducted an analysis on the subset of \textit{Journal of Neuroscience} papers with sub-disciplinary classifications (see Supplementary Information for further detail). The inclusion of these classifications into the model has little impact on either the extent of citation imbalance or the discrepancy in citation behavior across citing author gender (Figure \ref{fig:FigureS3}).

Second, because the citation distribution is long-tailed, it is of particular interest to determine whether the observed overcitation of men is driven primarily by a subset of highly cited papers \citep[e.g.,][]{abramo_contribution_2009}. To address this question, we separately quantified over/undercitation of gender categories within the top 50\% and bottom 50\% of the citation distribution (see Supplementary Information for further detail). While the papers in the top half of the distribution account for three-quarters of all citations in the data, the patterns of over/undercitation are highly similar for both sets (Figure \ref{fig:FigureS4}). This finding suggests that the observed imbalance is not driven by a `rich club' of papers authored by men.

\begin{figure*}[ht]
	\centering
	\includegraphics[width=1\linewidth]{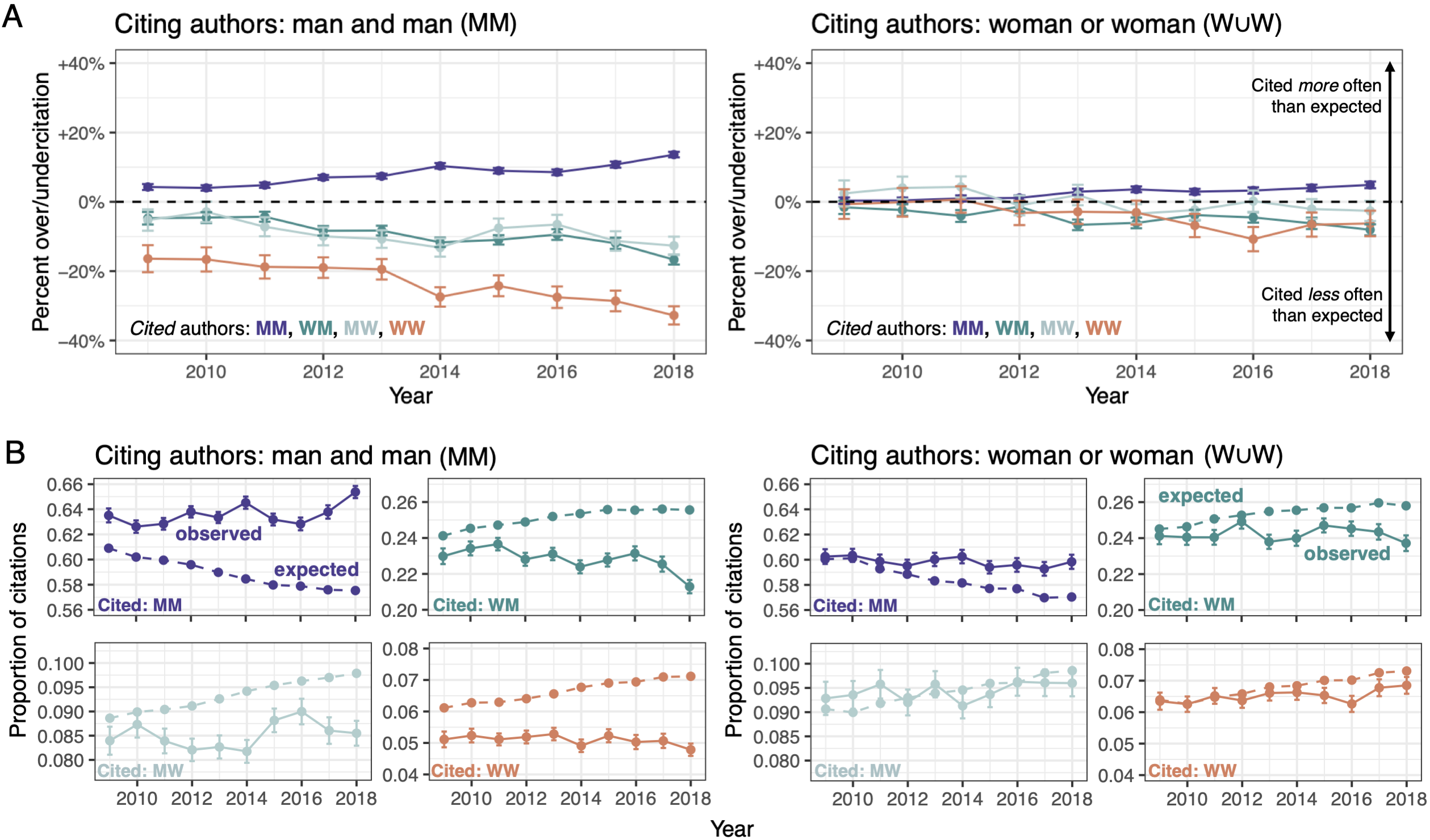}
	\caption{\textbf{Temporal trends in citation rates across cited and citing author gender.} \textit{(A)} Extent of over/undercitation across author gender categories over time, within MM (left) and W$\cup$W (right) reference lists. \textit{(B)} Observed (solid line) and expected (dotted line) citation proportions within MM reference lists (left) and W$\cup$W reference lists (right). Within each section, top left shows the observed/expected proportion of citations given to MM papers over time, top right shows WM papers, bottom left shows MW papers, and bottom right shows WW papers. The figure demonstrates relatively static observed proportions across groups, while expected proportions change to reflect increasing diversity within the field.}
	\label{fig:Figure4}
\end{figure*}

\subsection*{Temporal trends of citation imbalance}
In addition to the overall citation behavior, it was of interest to quantify the time-varying gender imbalance as the field has become more diverse over the years. As an intuitive measure of the overcitation of men in the literature, we specifically examined the absolute difference between the observed proportion of MM citations and the expected proportion of MM citations. We found that the gap between observed and expected proportions has been growing at a rate of roughly 0.41 percentage points per year (95\% CI = [0.34, 0.49], $p<0.001$). This finding suggests that citation practices are becoming less reflective of an increasingly diverse body of researchers; this is in contrast to the hypothesis that undercitation of women-led papers will be decreasing over time (Hypothesis 3).

Importantly, this growing gender gap does not simply reflect authors’ propensity to cite older literature from when the field was more men-dominated, as the expected proportions account for the publication year of the articles being cited. In other words, under this construction of expected proportions, both observed and expected citation rates for each gender category would remain constant if authors were to simply cite the same literature year after year. Thus, an expanding gap between observed and expected rates suggests that either observed proportions of MM citations are increasing (e.g., authors are citing more men than they used to), expected proportions of MM citations are decreasing (e.g., the newer literature is written by a more diverse field of authors), or some combination of both.

Similar to the overall tendency of MM teams to overcite other MM papers to a greater extent than W$\cup$W teams do, we found that the degree of overcitation has been increasing faster within MM reference lists than within W$\cup$W reference lists. Specifically, the absolute difference between the observed and expected proportions of MM citations is growing at a rate of 0.54 percentage points per year (95\% CI = [0.43, 0.63], $p<0.001$; Figure \ref{fig:Figure4}A, left) within MM reference lists, and it is growing at a rate of 0.29 percentage points per year (95\% CI = [0.17, 0.40], $p=0.023$; Figure \ref{fig:Figure4}A, right) within W$\cup$W reference lists. The fact that overcitation of MM papers is rising faster in MM reference lists than W$\cup$W reference lists ($p=0.014$) is related to the second aspect of Hypothesis 3, though the predicted temporal trend is flipped.

Further analysis revealed that the increasing overcitation of men in MM reference lists, and the moderately increasing overcitation of men in W$\cup$W reference lists, reflect relatively stable citation proportions for MM papers in the face of decreasing expected proportions over time (Figure \ref{fig:Figure4}B). Specifically, the proportion of MM papers within MM reference lists has been increasing slightly, at a rate of roughly 0.15 percentage points per year (95\% CI = [0.03, 0.26]). This proportion has not been clearly increasing or decreasing within W$\cup$W reference lists, changing with a rate of -0.08 percentage points per year (95\% CI = [-0.19, 0.04]). These findings demonstrate that although the rate at which scholars cite men has been relatively stable, this lack of change has led gender proportions within reference lists to be increasingly unrepresentative of the diversifying field.

\subsection*{The relationship between social networks and citation behavior}
Recent work has shown that researchers are more likely to work with other researchers of their own gender (i.e., homophily exists within co-authorship networks), and that such homophily in social networks can produce biased perceptions of the overall gender make-up of a network \citep{holman_researchers_2019,lee_homophily_2019}. Since homophily-driven perception biases in the overall gender make-up of the field could be a potential driver of the overcitation of men by men, and slight overcitation of women by women, we sought to estimate and isolate the relationships between authors’ social networks and their citation behavior. Because citations occur at the level of individual published papers, we developed two metrics to quantify gender imbalance within social networks at the paper level. Specifically, these measures consider the co-authorship network of the first and last authors of a given paper at the time of the paper’s publication. Thus, two papers written by the same authors may have different values for these measures, since the co-authorship network surrounding the authors may have changed over time. 

\begin{figure*}[ht]
\centering
\includegraphics[width=1\linewidth]{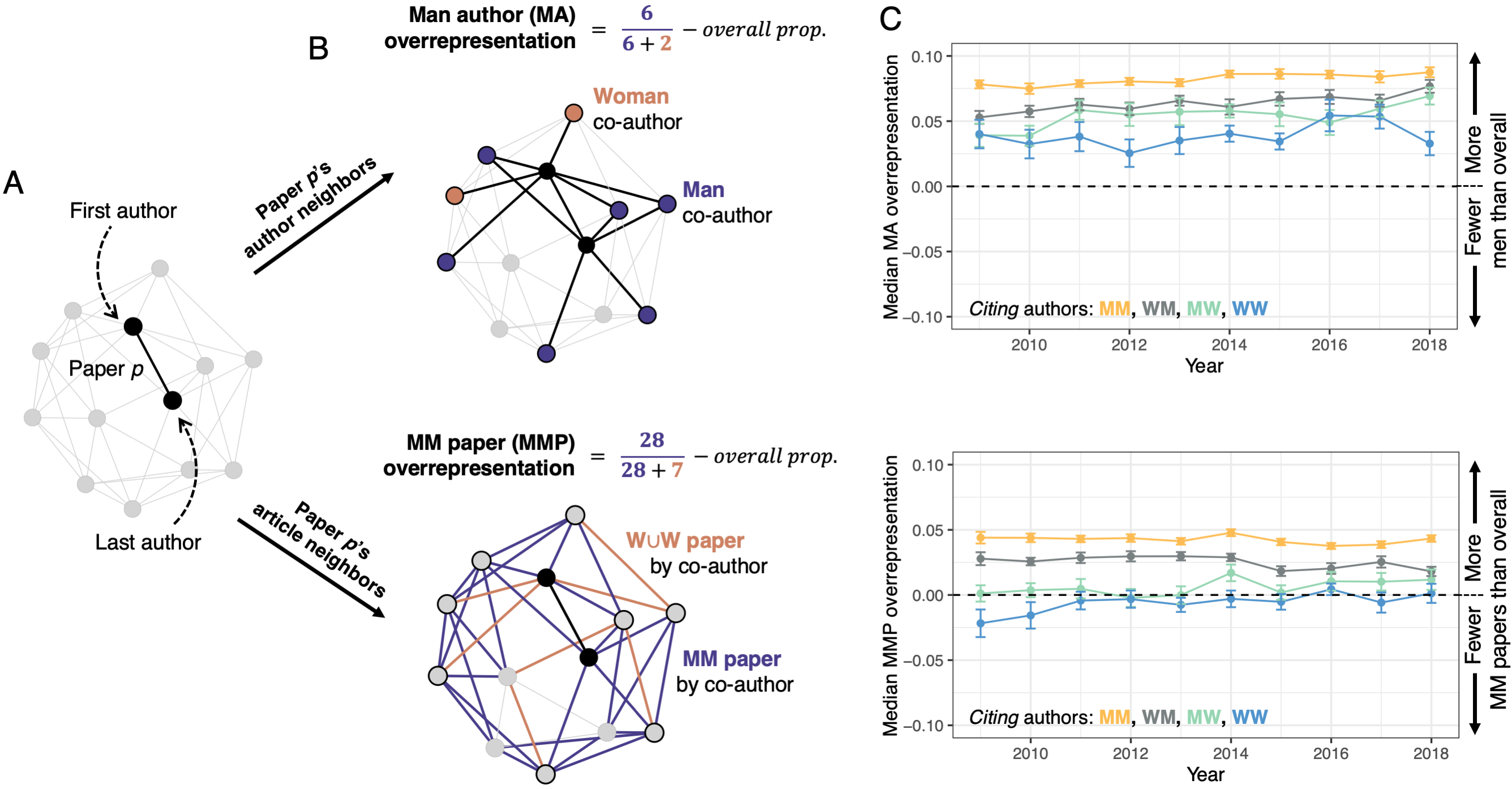}
\caption{\textbf{Visualization of co-authorship network composition measures.} \textit{(A)} Example region of a co-authorship network, where a specific article (edge) and the first and last author (nodes) are highlighted. \textit{(B)} Examples of the calculation of \textit{man author overrepresentation} ($MA_{or}$; top) and \textit{MM paper overrepresentation} ($MMP_{or}$; bottom) for the highlighted article. Here, $MA_{or}$ is the difference between the local proportion of men (purple nodes) and the overall proportion of men. The quantity $MMP_{or}$ is the difference between the local proportion of MM papers (purple edges) and the overall proportion of MM papers. \textit{(C)} Differences in the local network composition based on author gender. The panel shows that papers with more women tend to have less overrepresentation of men and man-led papers within their local networks.}
\label{fig:Figure5}
\end{figure*}

For a given paper, $p$, the first metric, which we refer to as man author overrepresentation, is defined as the difference between 1) the proportion of men within $p$'s author-neighborhood (defined as the union of researchers who had previously co-authored a paper with either the first or last author of $p$), and 2) the overall proportion of men within the network at the time of $p$’s publication. The second metric, which we refer to as MM paper overrepresentation, gives the difference between 1) the proportion of MM papers within $p$’s paper-neighborhood (defined as the union of all papers written by $p$'s first author, last author, or any of their previous co-authors), and 2) the overall proportion of MM papers within the network at the time of $p$’s publication. Visual examples of these two measures can be seen in Figure \ref{fig:Figure5}A-B (see Methods for further details).

We found that across groups, co-authorship networks tended to have more men than the field as a whole, but this overrepresentation of men within co-authorship networks was especially pronounced in the networks of MM teams. Specifically, the median MM team had roughly 8.2\% more men in their co-authorship network than the field’s base rate (95\% CI = [8.0, 8.4]), compared to the median WW team, which had roughly 3.8\% more men in their network than the field’s base rate (95\% CI = [3.3, 4.5]). Mixed gender teams fell in the middle, with their networks being comprised of around 6\% more men than the field’s base rate (WM = 6.4, 95\% CI = [6.1, 6.7]; MW = 5.7, 95\% CI = [5.2, 6.1]; Figure 5C). The overrepresentation of MM papers among those written by authors’ previous collaborators also differed based on citing authors’ gender. Yet in this case, MM papers were overrepresented relative to their overall proportion only within the social networks of MM teams (+4.2\%, 95\% CI = [4.1, 4.4]) and WM teams (+2.5\%, 95\% CI = [2.3, 2.8]). MM papers were roughly proportionally represented within networks of MW teams (+0.7\%, 95\% CI = [0.2, 1.0]) and were slightly underrepresented within networks of WW teams (-0.4\%, 95\% CI = [-0.8, -0.1]; Figure \ref{fig:Figure5}C).

\begin{figure*}[ht]
	\centering
	\includegraphics[width=1\linewidth]{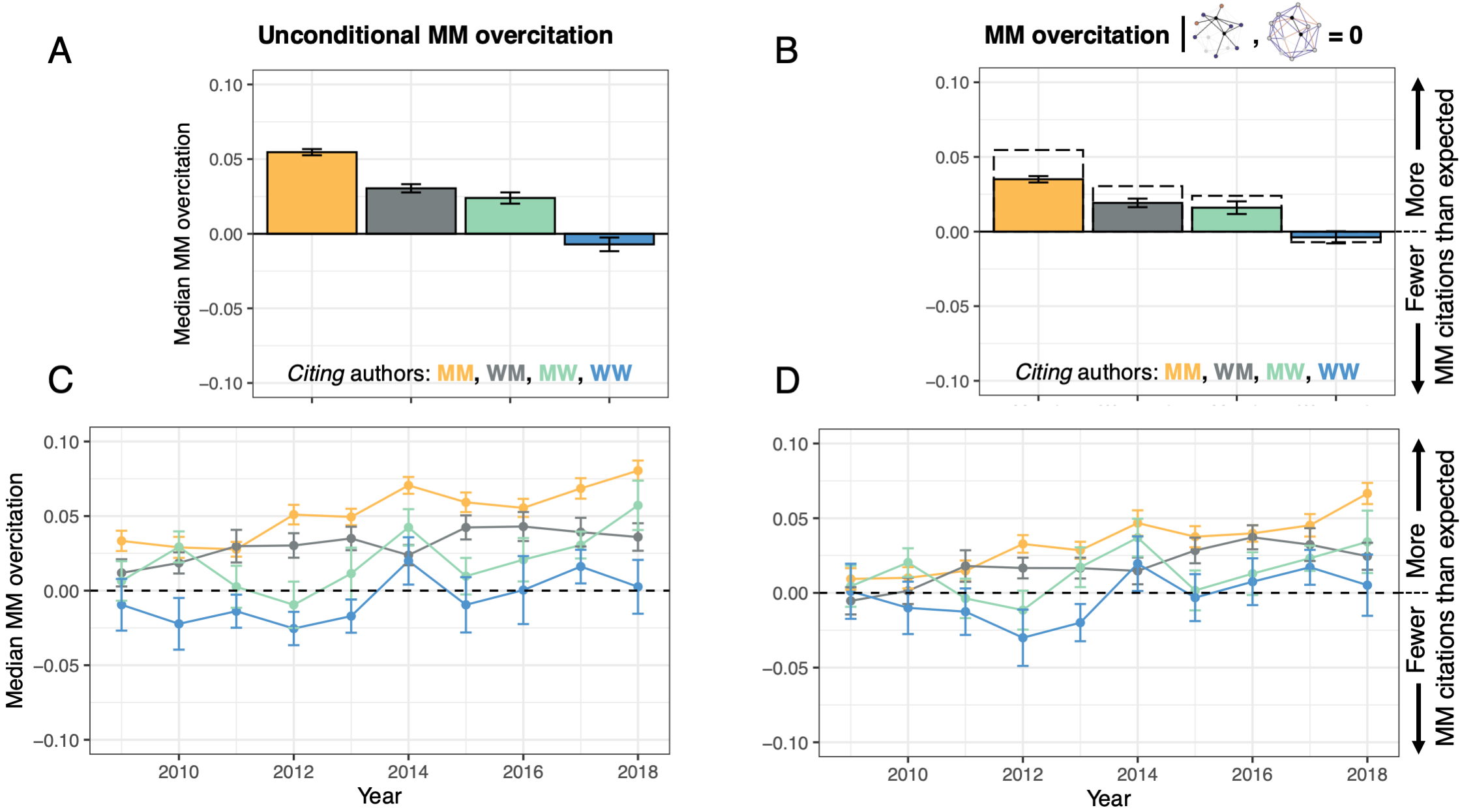}
	\caption{\textbf{Overcitation of MM papers before and after accounting for local network composition.} \textit{(A)} Overcitation of MM papers by author gender. The panel shows that MM, WM, and MW papers tend to overcite MM papers relative to expectation, while WW papers cite MM and W$\cup$W papers at roughly the expected rate. \textit{(B)} Overcitation of MM papers by author gender, after accounting for network effects. The panel shows that local network composition explains some of the group differences, but the general pattern remains. \textit{(C)} Overcitation of MM papers is increasing over time across groups. \textit{(D)} Overcitation is increasing over time across groups even after accounting for authors' social network effects.}
	\label{fig:Figure6}
\end{figure*}

Because gendered differences in social networks tended to follow similar patterns to gendered differences in citation behavior, it was of interest to determine the degree to which the composition of authors’ networks accounts for overcitation of men. For this analysis, we again utilized the absolute difference between the observed proportion of MM citations within a paper’s reference list and the expected proportion based on the characteristics of the cited papers. Without accounting for differences in authors’ social networks, we found that the median MM team overcites MM papers by roughly 5.5 percentage points (95\% CI = [5.1, 5.8], $p<0.001$), compared to 3.0 for WM teams (95\% CI = [2.6, 3.6], $p<0.001$), 2.4 for MW teams (95\% CI = [1.6, 2.9], $p=0.008$), and -0.7 for WW teams (95\% CI = [-1.7, 0.3], $p>0.99$; Figure \ref{fig:Figure6}A).

To estimate and account for the role of authors' social networks, we modeled papers' degree of MM overcitation as a function of author gender category, man author overrepresentation, and MM paper overrepresentation. Because the overcitation measure is bounded and skewed, we performed quantile regression to obtain estimates of the conditional median (see Methods for further details). Both MM paper overrepresentation and man author overrepresentation were independently associated with MM overcitation; a one percentage point increase in local overrepresentation of MM papers corresponded to a 0.24 percentage point increase in median MM overcitation (95\% CI = [0.21, 0.28], $p<0.001$), and a one percentage point increase in local overrepresentation of man authors corresponded to a 0.09 percentage point increase in median MM overcitation (95\% CI = [0.05, 0.12], $p<0.001$). These relationships are consistent after accounting for author seniority, though they appear to be slightly stronger among more senior teams (see Supplementary Information and Figure \ref{fig:FigureS5} for further detail). Thus, the data do support the hypothesis that there is a relationship between local co-authorship networks and citation behavior (Hypothesis 4).

However, after accounting for the degree of overrepresentation of both men and MM papers within authors’ social networks, differences in citation behavior remained across citing authors’ gender. Specifically, conditional on authors’ networks being representative of the field as a whole (i.e., local overrepresentation of men = 0 and local overrepresentation of MM papers = 0), the median MM team would still be expected to overcite MM papers by roughly 3.5 percentage points (95\% CI = [3.1, 3.9], $p<0.001$), compared to 1.9 for WM teams (95\% CI = [1.4, 2.5], $p=0.023$), 1.6 for MW teams (95\% CI = [0.7, 2.3], $p=0.18$), and -0.4 for WW teams (95\% CI = [-1.0, 0.7], $p>0.99$; Figure \ref{fig:Figure6}B). These results suggest that local homophily explains only part of the overcitation of men by other men. The findings also demonstrate that only WW papers tend to cite the expected proportion of MM papers both before and after accounting for network effects, while mixed or two-man teams tend to overcite MM papers in both cases.

Notably, accounting for network effects has almost no impact on the temporal trend of MM overcitation. Specifically, we find that the degree to which MM papers are overcited in reference lists has been increasing at an almost identical rate of $\sim$0.45 percentage points per year both before (0.44, 95\% CI = [0.34, 0.54]; Figure \ref{fig:Figure6}C) and after (0.46, 95\% CI = [0.33, 0.55]; Figure \ref{fig:Figure6}D) accounting for network measures. This trend suggests that although social networks are associated with the magnitude of MM overcitation, they are likely not a driver of reference lists being \textit{increasingly} unrepresentative over time. 

\section*{Discussion}
Like many scientific disciplines, the field of neuroscience currently faces many structural and social inequities, including marked gender imbalances \citep{schrouff_gender_2019}. While the task of addressing these imbalances often depends in part on people in positions of power (e.g., journal editors \citep{noauthor_promoting_2018}, grant reviewers and agencies \citep{bornmann_gender_2007,jagsi_sex_2009,van_der_lee_gender_2015}, department chairs \citep{nielsen_limits_2016,van_den_brink_scouting_2011,de_paola_gender_2015}, and presidents of scientific societies \citep{joels_tale_2014}), many imbalances are caused and perpetuated by researchers at all levels. One example is imbalance within citation practices \citep{ferber_gender_2011,maliniak_gender_2013}. Although the usefulness of citations as a measure of scientific value is tenuous \citep{aksnes_citations_2019}, the engagement that they represent can affect how central to a field scholars are viewed to be by their peers \citep{ferber_gender_2011}. This impact on perception can then have downstream effects on conference invitations, grant and fellowship awards, tenure and promotion, inclusion in syllabi, and even student evaluations. As a result, understanding and eliminating gender bias in citation practices is vital for addressing gender imbalances in a field.

In this study, we sought to determine whether there is evidence of gender bias in neuroscience citations, and whether that bias itself differs based on the gender of the citing authors. We indeed found evidence that neuroscience reference lists tend to include more papers with men as first and last author than would be expected if gender was not a factor. Importantly, this overcitation of men and undercitation of women is driven largely by the citation practices of men. Specifically, papers with men as first and last author overcite other man/man papers by 8\% relative to the expected proportion, undercite woman/man papers by 9\%, undercite man/woman papers by 9\%, and undercite woman/woman papers by 23\%. Papers with women in one or both primary authorship positions overcite man/man papers by 2.5\%, undercite woman/man papers by 4.5\%, undercite man/woman papers by 0.1\%, and undercite woman/woman papers by 4\%. These results are consistent with results from other fields that show that men are less likely to cite work by women \citep{ferber_gender_2011,maliniak_gender_2013,dion_gendered_2018}.

Gender inequity in general — and, one might therefore argue, gender inequity in citational practices — is understood to result from both systemic bias and individual bias. Systemic bias, also known as institutional bias, refers to discriminatory values, practices, and mechanisms that function at the intergroup level in the domain of social institutions \citep{henry_institutional}. Individual bias may be either explicit or implicit. Explicit bias is consciously held or expressed prejudice against a particular group or an individual of that group, resulting in material, psychological, or physical harms \citep{clarke_explicit_nodate}. Implicit bias, on the other hand, is a set of subconsciously harbored discriminatory attitudes against a particular group or an individual of that group, which can result in prejudicial speech and social behaviors \citep{brownstein_implicit_2019,greenwald_implicit_1995}. Implicit bias is traceable in individual attitudes (and resulting actions) relative to something as concrete as physical appearance and as abstract as a mere name. Indeed, implicit bias with respect to names has been shown in studies of race-based \citep{conaway_implicit_2015,bertrand_are_2004} and gender \citep{paludi_whats_1985,macnell_whats_2015} discrimination. The undercitation of women in neuroscience papers, therefore, may be due to systemic gender bias or to explicit or implicit individual bias relative either to the known gender of an author or to an author’s gendered name. Our analysis thus extends and contributes to existing literature on bias, gender inequity, and citational practices.  

Recent work has also shown that homophily in social networks (i.e., an increased likelihood of being connected to people of the same gender) can lead to biases in individuals’ perceptions of the overall proportion of men and women \citep{lee_homophily_2019}. Prior work has indeed found evidence for homophily in scientific collaboration \citep{holman_researchers_2019,jadidi_gender_2018}. In this context, such an effect could potentially drive overall imbalance and differential citation behavior between men-led and women-led teams. To determine the degree to which homophily in researchers’ social networks explains the prevalence of men overciting other men, we developed a co-authorship network and measured the degree to which men (and man-led papers) were overrepresented in authors’ local neighborhoods. We find that while these features are indeed associated with authors’ overcitation of men, man/man papers are still overcited in the reference lists of other man/man papers, and to a lesser extent in reference lists of woman/man and man/woman papers, even after accounting for network effects. The reference lists of woman/woman papers are the only group that tend to cite man/man papers at the expected rate. These results, while consistent with the presence of some homophily effect, show that men tend to overcite other men even if their social networks are representative of the field.

There are several possible mechanisms that could explain the remaining difference in citation behavior between man-led teams and woman-led teams. One obvious explanation is greater conscious or unconscious bias among men, which could lead them to evaluate woman-led work more harshly and thus be more hesitant to cite such work. This explanation would be consistent with studies that have shown evaluative bias in the realms of graduate admissions \citep{posselt_inside_2016}, faculty hiring \citep{moss-racusin_science_2012}, grant funding \citep{van_der_lee_gender_2015}, and promotion \citep{de_paola_gender_2015}.

Other explanations, like the overrepresentation of men in course syllabi \citep{colgan_gender_2017} and in conference speaking roles \citep{schrouff_gender_2019}, could partly explain the difference between groups (e.g., women may take more courses taught by women, who discuss and assign more work by women). Yet mechanisms like these would likely be more consistent with an overall overcitation of men that does not differ based on the gender of the citing authors. In that case, the fact that teams with more women display less gender citation bias could be explained by their conscious efforts to seek out and cite work by other women. If this is the case, it is plausible that women’s attempts to address gender imbalance could make their citation practices more representative of their fields, while men’s indifference or lack of awareness could lead to the propagation of imbalances present in syllabi and conferences.

Regardless of the mechanisms that drive these imbalances, greater awareness of existing (and persisting, and even increasing) imbalances in citation practices is likely an important step in heightening researchers’ willingness to address these issues. Recent work has laid out guidelines for responsible citation practices \citep{penders_ten_2018}, which include consideration of gender imbalance. There also exist tools to probabilistically measure the proportion of women and men within course syllabi and reference lists \citep{sumner_gender_2018}. Various organizations also provide information that can assist researchers in creating representative reference lists. These include BiasWatchNeuro (biaswatchneuro.com), which publishes base rates for different subfields within neuroscience, and Women in Neuroscience (winrepo.org) and Anne’s List (anneslist.net), which contain detailed, searchable databases of women in neuroscience and their areas of expertise. Neuroscience might also consider having a reference list that represents additional axes of marginalization (e.g. gender, race, class, sexuality, disability, citizenship, etc.), given both the intersectional discrimination of women in the academy \citep{gutierrez_y_muhs_presumed_2012} and the aspiration to address social inequities in the field more broadly. The American Philosophical Association’s UP Directory \citep{noauthor_up_nodate} provides a potential reference point for this type of inclusive list. 

Addressing the identified imbalances will require researchers, particularly men, to make use of available resources and engage in more thoughtful citation practices. Efforts can also be made by journal editors and reviewers to inform authors of these issues and encourage transparency within manuscripts. This paper, for example, includes a \textit{Citation gender diversity statement} that describes the gender make-up of its reference list. Educating graduate students about citation practices will also be vital, and such discussions could potentially be included in the NIH and NSF's ``responsible conduct of research (RCR)" requirements.

Beyond a growing individual and collective thoughtfulness, paired with ad-hoc efforts, to achieve gender balance in neuroscience reference lists, the ethics of citation practices remain to be further defined. Righting social inequities may be accomplished on a number of different models. On the distributive model, for example, justice refers to the morally proper distribution of social goods and resources or, in this case, citations \citep{rawls_theory_1999}. Exactly how that distribution ought to be circumscribed, however, remains in question. On the equality-based distributive model, citations ought to be allocated to all authors equally; while on the equity-based distributive model, citations ought to be allocated to authors differentially based upon select factors, which may include merit, need, or authority \citep{lamont_distributive_2017,olsaretti_idea_2018}. The distributive paradigm on the whole, however, is limited insofar as it emphasizes commodity parity across economies of exchange over differential responsibility for histories and structures of inequality \citep{young_justice_2011,ahmed_being_2012}. Difference models, by contrast, recommend acts of reparative justice \citep{walker_what_2010}, which might include affirmative action \citep{anderson_imperative_2010} in citational practices, institutional reform to support citation parity, and disciplinary redress of gender bias more broadly.

Distributive and difference-based models raise a series of important questions for citation ethics in neuroscience. Should gender balance in citation practices reflect random distributions or distributions tuned to relevant features (and, if so, which features)? Are such distributional structures sufficient either to correct for a history of underrepresentation or to secure a future of equitable representation? Given the lassitude with which social change occurs—and the worsening of gender imbalance in citation practices in neuroscience overall—is it justifiable for some research teams to overcite papers produced by women-led teams wherever possible? Might the effects of systemic gender bias on undercitation practices be counteracted by the significant employment of women in field-specific decision-making bodies, reforming checkpoints, and professional activism? And, given the function of implicit bias and its capacity for correction via experience, should researchers of all genders commit to collaborating more robustly with women and other gender minorities?

Overall, the work of citation is an important element in the research ethics of any field. Insofar as citation patterns today have inescapable effects on the future of neuroscience, citational practices in the field warrant more serious attention.

\subsection*{Limitations and future work}
This work is subject to several limitations. First, this study focuses on citing and cited articles published in five top neuroscience journals. Although this focus has the benefit of reducing the confounding effects of journal prestige, it also likely limits the degree to which the results can be generalized to neuroscience as a whole. Future work would be needed to understand the role of more specialized journals. Additionally, this study does not address the potential effects of authors' institutional prestige \citep{way_productivity_2019}. As a result, a combination of gender imbalance in hiring and prestige-based citation behavior could introduce bias in the results. Future work could attempt to quantify and isolate the effect of departmental prestige in this context. 

Importantly, the methods used for gender determination are limited to binary man/woman gender assignments. This study design, therefore, is not well accommodated to intersex, transgender, and/or non-binary identities, and incorrectly assumes that all authors in the dataset can be placed into one of two categories. Ideally, future work will be able to move beyond the gender binary, potentially by applying methods that utilize pronouns or other forms of self-identification. The current study is also limited to investigating biases solely along gender lines. Future work could extend these types of analyses to examine biases along, for example, race or ethnicity, as well as their intersection with gender.

Efforts to expand the methodological scope of this study could build on the analysis of collaboration networks. While the present study investigates the role of network structure in citation behavior, prior work has found that the gender \citep{jadidi_gender_2018,yang_networks_2019}, ethnic \citep{alshebli_preeminence_2018}, and international \citep{uhly_gendered_2015,zippel_women_2017} composition of scholars' collaboration networks can be related to scientific impact and career success. Such effects in this data could provide additional information about mechanisms and potential points of intervention. Additionally, future work could incorporate longitudinal within-author analyses of citation behavior. Such analyses may provide a better understanding of how authors' practices change over time and are impacted by the gender of their co-authors; this knowledge could facilitate more individualized recommendations.

{\normalsize \section*{Methods}}
\begin{small}
\subsection*{Data collection}
We drew data for this study from the Thomson Reuters’ Web of Science (WoS) database. The Web of Science database indexes neuroscience journals according to the Science Citation Index Expanded, and we selected the neuroscience journals with the five highest Eigenfactor scores for study. Eigenfactor scores give a count of incoming citations, where citations are weighted by the impact of the citing journal. Therefore, this measure roughly mimics the classic version of Google page rank, and attempts to characterize the influence a journal has within its field \citep{bergstrom_eigenfactortm_2008,bergstrom_eigenfactor:_2007}. The journals selected were \textit{Brain}, \textit{Journal of Neuroscience}, \textit{Nature Neuroscience}, \textit{NeuroImage}, and \textit{Neuron}.

All articles published between 1995 and 2018 were downloaded, and articles classified as articles, review articles, or proceedings papers that were labeled with a Digital Object Identifier (DOI) were included in the analyses. The data downloaded from WoS included papers’ author names, reference lists, publication dates, and DOI, and we obtained information on each paper’s referencing behavior by matching DOIs contained within a reference list to DOIs of papers included in the dataset.

Although authors’ last names were included for all papers, authors’ first names were only regularly included in the data for papers published after 2006. For all papers published in or before 2006, we searched for author first names using Crossref’s API. When first names were not available on Crossref, we searched for them on the journals’ webpage for the given article. To minimize the number of papers for which we only had access to authors’ initials, to remove self-citations, and to develop a co-authorship network, we implemented a name disambiguation algorithm. 

\subsection*{Author name disambiguation}
To minimize missing data, allow for name gender assignment, and allow for author matching across papers, we implemented an algorithm to disambiguate authors for whom different versions of their given name or initials were available across papers. We began by separating first and last names according to the method used by the given source (e.g., WoS typically used “last, first; last, first”). We then identified cases in which only initials were available after the previously described searching steps by marking authors for whom the first name entry contained only uppercase letters (as we found that many initials-only entries did not contain periods). 

For each case, we collected all other entries that contained the same first/middle initials and the same last name. If only one unique first/middle name matched the initials of the given entry, or if distinct matches were all variants of the same name, we assigned that name to the initials. If there were multiple names in the dataset that fit the initial/last name combination of the given entry, then we did not assign a name to the initials. For example, if an entry listed an author as R. J. Dolan, and we found matches under Ray J. Dolan and Raymond J. Dolan, we would replace the R. J. Dolan entry with the more common completed variant. If, instead, we found matches under Ray J. Dolan and Rebecca J. Dolan, we would not assign a name to the original R. J. Dolan entry.

Next we matched different name variants for the sake of tracking individual authors across their papers. To find and connect variants, we searched for instances of author entries with matching last names and either the same first name or first names that were listed as being commonly used nicknames according to the Secure Open Enterprise Master Patient Index \citep{toth_soempi:_2014}. If there were no matches that fit that description, the name was retained. If there was one match that occurred more commonly, the less common variant was changed to the more common variant. If there were multiple matches that did not have any conflicting initials (some having a middle initial and others not having one was not considered conflicting), then less common variants were changed to the more common variant. If there were multiple matches that did have conflicting initials (e.g., Ray Dolan being matched to both Raymond S. Dolan and Raymond J. Dolan), then the target name was not changed.

There are three primary ways that incorrect author disambiguation could impact the results presented in this study. First, inability to link initials to an author's first name would yield missing data for papers that only included the author's initials. These papers would then not be included in the analyses as either \textit{cited} or \textit{citing} papers. Second, inability to link two versions of an author's name (e.g., Ray Dolan and Raymond Dolan) would lead to the inclusion of some self-citations into the author's analyzed reference lists. This could lead to slightly inflated rates of authors citing other authors of the same gender, though sensitivity analyses suggest that this effect is essentially non-existent in the present data (see Supplementary Information and \ref{tab:tabdisam}). Third, incorrectly linking Author A to Author B would lead to the unnecessary removal of some citations (i.e., any of Author B's references to Author A's work would be removed as self-citations). Though this is likely a rare occurrence, its presence would lead to slightly decreased rates of authors citing other authors of the same gender.

\subsection*{Author gender determination}
For all authors with available first names, we carried out the process of gender determination in two steps. First, we used the Social Security Administration (SSA) database as implemented in the `gender' R package \citep{blevins_jane_2015}, which returns the proportion of matching baby names given to infants assigned female or male at birth between the years 1932 and 2012. The assignment of “woman” was given to names with greater than a 0.7 probability of being assigned female at birth, and the assignment of “man” was given to names with greater than a 0.7 probability of being assigned male at birth. Because this database is primarily useful for United States-based authors, we used Gender API as a secondary source for any names that had a probability between 0.3 and 0.7 or were not present in the SSA database. To determine the extent of potential gender mislabeling, we conducted a manual study on a sample of 200 authors. The relative accuracy of the automated determination procedure at the level of both individual authors (Accuracy $\approx$ 0.96; Table \ref{tab:tabvalaut}) and article gender categories (Accuracy $\approx$ 0.92; Table \ref{tab:tabvalpap}) can be seen in the Supplementary Information. Because errors in gender determination would break the links between citation behavior and author gender, any incorrect estimation in the present data likely biases the results towards the null.

\subsection*{Removal of self-citations}
For this study, self-citations were removed from all analyses of gendered citation behavior. Though self-citations themselves have been found to have relevant gendered properties \citep{king_men_2017}, their removal in this study allows us to isolate more comparable external citation behaviors of men and women in the field. In the Supplementary Information, we further explore the role of self-citations in this data. Specifically, Figure \ref{fig:FigureS1} shows the impact of including self-citations on the main results, and Table \ref{tab:tabself} shows the relative prevalence of self-citations across author genders.

For the primary analyses, we defined self-citations as papers for which either the cited first or last author was the first or last author on the citing paper. While this is a somewhat restrictive definition, it is the only type of self-citation for which the author gender of the cited paper is necessarily determined by the author gender of the citing paper. In the Supplementary Information, we demonstrate that using broader definitions of self-citation has little to no impact on the results. Specifically, Table \ref{tab:tabentlist} shows the results when the entire author list of the \textit{citing} paper was considered in the definition of self-citations, and Table \ref{tab:tabcoauth} shows the results when the entire author list of the \textit{cited} paper was considered in the definition of self-citations.

\subsection*{Statistical analysis}
Many analyses conducted in this study rely on comparisons between observed citation behavior and the rates at which MM, WM, MW, and WW papers would be expected to appear in reference lists if gender was irrelevant. To obtain expected rates that account for various characteristics that may be associated with gender, we fit a generalized additive model (GAM) on the multinomial outcome \{MM, WM, MW, WW\}, in which the model's features were 1) month and year of publication, 2) combined number of publications by the first and last authors, 3) number of total authors on the paper, 4) the journal in which it was published, and 5) whether it was a review paper. When this model is then applied to each paper, it yields a set of probabilities that the paper belongs to the MM, WM, MW, and WW categories, respectively. Importantly, this model does not predict the number of citations given to individual papers. Instead, it facilitates the calculation of the rates at which different gender categories would be expected to appear in reference lists if author gender was independent of citation rates, conditional on the other characteristics in the model. The GAM was fit using the `mgcv' package in R \citep{mgcv}, using penalized thin plate regression splines for estimating smooth terms of publication date, author experience, and team size.

Estimates in this study are presented with either a confidence interval, a \textit{p}-value, or both. Confidence intervals in this study were calculated by bootstrapping citing papers (i.e., randomly sampling citing papers with replacement). As opposed to bootstrapping individual instances of citations, this method maintains the dependence structure of the clusters of cited articles within citing articles. The null model used to obtain \textit{p}-values was derived from the randomization of \textit{cited} papers' author gender categories. Randomization was carried out by probabilistically drawing new gender categories for each paper according to their GAM-estimated gender probabilities. Randomly sampling gender categories for each paper therefore produces a null model in which cited author gender is conditionally independent of citation rates and citing author gender (conditional on the characteristics included in the GAM model), while the structure of the citation graph and the long-tailed nature of the citation distribution are both preserved. 10,000 randomizations were carried out to calculate \textit{p}-values (see Table \ref{tab:tabnull} for means and standard errors of estimated quantities across randomizations, and Figure \ref{fig:FigureS2} for a visualization of the null distributions). Because significance is assessed for multiple primary comparisons, all presented \textit{p}-values were corrected according to the Holm-Bonferroni method \citep{holm_simple_1979}.

In the following sections, we describe the formal statistical analysis that we used to address the four distinct hypotheses. In each subsection we state the hypothesis first, followed by the analysis used to test it. All hypotheses were tested for the set of articles published between 2009 and 2018. We decided to specifically consider reference lists from the past 10 years to ensure that estimates of over/undercitation reflected current behavior, and were not a result of aggregating over disparate eras of neuroscience research.

\subsubsection*{Hypothesis 1: The overall citation rate of women-led papers will be lower than expected given papers’ relevant characteristics}
To test this hypothesis, we first estimated the expected number of citations given to each author gender category. We calculated this expectation by summing over the GAM-estimated probabilities for all papers contained within the reference lists of citing papers. These totals, therefore, reflect the expected number of citations given to MM, WM, MW, and WW papers if author gender was conditionally independent of citation behavior, given the paper characteristics included in the model described above.

To calculate the observed number of citations given to each group, we simply summed over the \{MM, WM, MW, WW\} dummy variable for all of the papers contained within the reference lists of papers published between 2009 and 2018. These values were compared by calculating the percent difference from expectation for each author gender group. For example, for WW papers, this percent change in citation would be defined as,

$$
\Delta_{WW}=\frac{obs_{WW}-exp_{WW}}{exp_{WW}},
$$

\noindent where $obs_{WW}$ is the number of citations given to WW papers between 2009 and 2018, and $exp_{WW}$ is the expected number of citations given to WW papers between 2009 and 2018. 

Notably, performing the summation over all citations results in the upweighting of articles with many citations, and the downweighting of articles with few citations. This approach helps to improve the stability of the estimates, but could potentially be sensitive high-influence observations. To determine the impact of this decision, we conducted sensitivity analyses in which we used the mean of article-level over/undercitation, and found little difference between the two estimation strategies (see Supplementary Information and Table \ref{tab:tabsec} for further details). 

\subsubsection*{Hypothesis 2: The undercitation of women-led papers will occur to a greater extent within men-led reference lists}
To test this hypothesis, we used very similar metrics to those described in the previous section. The primary difference is that instead of calculating the observed and expected citations by summing over the citations within all reference lists between 2009 and 2018, in this section we performed those summations separately for reference lists in papers with men as first and last author (MM papers) and papers with women as first or last author (W$\cup$W papers). For example, to estimate the over/undercitation of WW papers within the reference lists of MM papers, we define,

$$
\Delta^{(MM)}_{WW}=\frac{obs^{(MM)}_{WW}-exp^{(MM)}_{WW}}{exp^{(MM)}_{WW}},
$$

\noindent where $obs^{(MM)}_{WW}$ is the total number of citations given to WW papers within MM reference lists, and $exp^{(MM)}_{WW}$ is the expected number of citations given to WW papers within MM reference lists.

\subsubsection*{Hypothesis 3: Undercitation of women-led papers will be decreasing over time, but at a slower rate within men-led reference lists}
As there are four separate measures representing over or undercitation of each author group, we calculated change in the overcitation of men over time using the simple measure of the absolute difference between the observed proportion of MM papers cited and the expected proportion of MM papers cited. This measure of change is given by, 

$$
\delta_{MM,year}=\frac{obs_{MM,year}-exp_{MM,year}}{obs_{year}},
$$

\noindent where $obs_{year}$ is the total number of citations within a given year, $obs_{MM,year}$ is the number of citations given to MM papers in a specific year, and $exp_{MM,year}$ is the expected number of citations given to MM papers in a specific year. The change in the overcitation of men over time is estimated using a linear regression of $\delta_{MM,year}$ on year; the confidence interval of this estimate is obtained using the article bootstrap procedure; and significance is assessed using the graph-preserving null model.

Similarly, to estimate the change in overcitation of MM papers separately within MM reference lists and W$\cup$W reference lists, we defined group-specific measures of yearly overcitation. For example, overcitation of MM papers within MM reference lists for a specific year would be given by,

$$
\delta^{(MM)}_{MM,year}=\frac{obs^{(MM)}_{MM,year}-exp^{(MM)}_{MM,year}}{obs^{(MM)}_{year}},
$$

\noindent where $obs^{(MM)}_{year}$ is the total number of citations within MM reference lists in a specific year, $obs^{(MM)}_{MM,year}$ is the number of citations given to MM papers within MM reference lists in a specific year, and $exp^{(MM)}_{MM,year}$ is the expected number of citations given to MM papers within MM reference lists in a specific year.

\subsubsection*{Hypothesis 4: Differences in undercitation between men-led and women-led reference lists will be partly explained by the structure of authors’ social networks}
To test this hypothesis, we developed a temporal co-authorship network in which nodes were individual authors (only authors who appeared as first or last author in at least one paper in the dataset were included), and binary edges represented the fact that two authors had appeared on at least one paper together prior to a given date. It is of interest in this section to estimate the relationship between authors’ local network composition and their citation behavior. Because citation behavior occurs at the level of a reference list within a specific paper with both a first and a last author (rather than at the level of a single node, or author), we sought to define two measures of local network composition at the paper level. For the purposes of these analyses, we consider a paper to be the set $\{a_f,a_l,m\}$, where $a_f$ is the first author, $a_l$ is the last author, and $m$ is the month of publication. We then define a paper’s local neighborhood of authors, $N_a^p$, to be the authors that are connected by shared publication to either $a_f$ or $a_l$ prior to month $m$. We also define a paper’s local neighborhood of papers, $N_p^p$, to be the union of all papers authored by anyone within $N_a^p$ prior to month $m$.

The two measures of local network composition are man author overrepresentation and MM paper overrepresentation. We define man author overrepresentation as the difference between the proportion of men within a paper’s local author neighborhood, $N_a^p$, and that of the overall network. For paper $p$, this measure is therefore given by,

$$
MA_{or}(p)=\pi_{M,N_a^p}-\pi_M,
$$

\noindent where $\pi_M$ is the proportion of men in the full co-authorship network, and $\pi_{M,N_a^p}$ is the proportion of men within paper $p$’s local author neighborhood. Similarly, we define MM paper overrepresentation as the difference between the proportion of MM articles within a paper’s local paper neighborhood, $N_p^p$, and that of the overall network. For paper $p$, this measure is therefore given by,

$$
MMP_{or}(p)=\pi_{MM,N_p^p}-\pi_{MM},
$$

\noindent where $\pi_{MM}$ is the overall proportion of MM articles within the data, and $\pi_{MM,N_p^p}$ is the proportion of MM articles within paper $p$’s local paper neighborhood.

To estimate the relationship between these metrics and the degree of overcitation of men within reference lists, we defined a paper-level measure of the absolute difference between the observed and expected proportion of MM papers. Similar to the previously described $\delta^{(MM)}_{MM,year}$ measure that quantified the overcitation of MM papers within all MM reference lists from a given year, here we define a measure of overcitation within an individual paper, $p$. It is given by,

$$
\delta^{(p)}_{MM}=\frac{obs^{(p)}_{MM}-exp^{(p)}_{MM}}{obs^{(p)}},
$$

\noindent where $obs^{(p)}_{MM}$ is the number of MM citations within paper $p$’s reference list, $exp^{(p)}_{MM}$ is the expected number of MM citations within paper $p$’s reference list based on the GAM-estimated assignment probabilities of each cited paper, and $obs^{(p)}$ is the total number of candidate citations within paper $p$’s reference list.

The relationships between $\delta^{(p)}_{MM}$, $MMP_{or}(p)$, $MA_{or}(p)$, and $\{MM, WM, MW, WW\}$ are estimated using weighted quantile regression, with the MM overcitation metric, $\delta^{(p)}_{MM}$, as the outcome. We performed quantile regression because of the bounded and skewed nature of the $\delta^{(p)}_{MM}$ measure, but the results of a sensitivity analysis using linear regression can be found in Table \ref{tab:tablin}. We define the weights to be equal to the number of candidate citations within a given paper’s reference list; this choice gives higher weight to papers for which the outcome is more stable. Results from an unweighted model can be found in Table \ref{tab:tabsec}. We also take the $\tau$ value of the quantile regression formula to be 0.5, resulting in a model fit to the median of the outcomes. Confidence intervals are again obtained by the article bootstrap method, and significance is assessed using the graph-preserving null model.
\end{small}


\vspace{.3cm}
\begin{acknowledgements}
\noindent We thank David Lydon-Staley and Dale Zhou for constructive comments on an earlier version of this manuscript. RTS would like to acknowledge support from the National Institute of Neurological Disorders and Stroke (R01 NS085211 \& R01 NS060910). DSB would like to acknowledge support from the John D. and Catherine T. MacArthur Foundation, the Alfred P. Sloan Foundation, and the National Science Foundation CAREER (PHY-1554488). The content is solely the responsibility of the authors and does not necessarily represent the official views of any of the funding agencies.
\end{acknowledgements}
\vspace{-.25cm}

\begin{refdiverse}
\noindent The gender balance of papers cited within this work was quantified using a combination of automated gender-api.com estimation and manual gender determination from authors' publicly available pronouns. Among the 71 cited works that had named authors, 34\% ($n = 24$) were MM, 13\% ($n = 9$) were WM, 11\% ($n = 8$) were MW, and 42\% ($n = 30$) were WW.
\end{refdiverse}

\begin{datacode}
\noindent Data that support the findings of this study, and code that can be used to reproduce estimates and figures, have been deposited in an Open Science Framework repository and can be accessed at https://bit.ly/2O2xEFi.
\end{datacode}

\begin{authorcont}
\noindent Conceptualization, JDD, KAL, RTS, and DSB; Methodology, JDD, KAL, EGT, RTS, and DSB; Data Curation, JDD; Formal Analysis, JDD; Writing – Original Draft, JDD, PZ, and DSB; Writing – Review \& Editing, JDD, KAL, EGT, PZ, RTS, and DSB; Funding Acquisition, RTS and DSB; Supervision, RTS and DSB.
\end{authorcont}

\onecolumn
\newpage

\appendix
\newcommand{\hbAppendixPrefix}{S}
\renewcommand{\thefigure}{\hbAppendixPrefix\arabic{figure}}
\setcounter{figure}{0}
\renewcommand{\thetable}{\hbAppendixPrefix\arabic{table}} 
\setcounter{table}{0}

\noindent\LARGE{\textbf{Supplementary Information for ``The extent and drivers of gender imbalance in neuroscience reference lists”}}\\

\noindent\normalsize{Jordan D. Dworkin$^1$, Kristin A. Linn$^1$, Erin G. Teich$^2$, Perry Zurn$^3$, Russell T. Shinohara$^1$ \& Danielle S. Bassett$^{2,4,5,6,7,8}$}\\

\noindent\small{$^1$Penn Statistics in Imaging and Visualization Center, Department of Biostatistics, Epidemiology, and Informatics,Perelman School of Medicine, University of Pennsylvania, Philadelphia, PA, USA\\$^2$Department of Bioengineering, School of Engineering and Applied Science, University of Pennsylvania, Philadelphia, PA, USA\\$^3$Department of Philosophy and Religion, American University, Washington DC, USA\\$^4$Department of Physics \& Astronomy, College of Arts and Sciences, University of Pennsylvania, Philadelphia, PA, USA\\$^5$Department of Electrical \& Systems Engineering, School of Engineering and Applied Science, University of Pennsylvania, Philadelphia, PA, USA\\$^6$Department of Neurology, Perelman School of Medicine, University of Pennsylvania, Philadelphia, PA, USA\\$^7$Department of Psychiatry, Perelman School of Medicine, University of Pennsylvania, Philadelphia, PA, USA\\$^8$Santa Fe Institute, Santa Fe, NM, USA}

\clearpage

\subsubsection*{Validation of gender determination}
Because of the degree to which the results presented in this work rely on accurate estimation of author genders, it was vital to understand the performance of the gender determination methods in this data. To test the performance of the gender assignments, a random sample of 100 papers were drawn from the full data, yielding 200 first and last author names. For each of these 200 authors, manual gender determination was conducted by assigning inferred gender via publicly available pronouns (e.g., on lab websites or social media pages). Although authors who use `she’/`her’(`he’/`him’) may identify as any number of genders, their gender is most likely to be `woman’(`man’) and least likely to be `man’(`woman’). For authors without publicly available pronouns, a gender determination was conducted by assigning inferred gender via publicly available photographs. Although authors whose gender expression most approximates `woman’(`man’) may identify as any number of genders, their gender is most likely to be `woman’(`man’) and least likely to be `man’(`woman’). Authors for whom photographs were ambiguous or unavailable, or whose names could not be reliably matched to one scholar, were marked as `unknown.’ Relative to this manual assignment, the accuracy of the automated gender assessment method can be seen at the author level in Table \ref{tab:tabvalaut}, and the article level in Table \ref{tab:tabvalpap}.

\begin{table*}[ht]
\centering
\small
\begin{tabular}{rrrr}

   & \multicolumn{3}{c}{Manual assessment} \\
  \cmidrule(r{5pt}){2-4}
 Automated assessment & Man & Woman & Unknown\\
 \midrule
 Man & 140 & 5 & 7 \\
 Woman & 2 & 42 & 4 \\
   \bottomrule
\end{tabular}
\captionsetup{width=1\linewidth}
\caption{\textbf{Accuracy of automated gender determination technique in a random sample of 200 authors}.}
\label{tab:tabvalaut}
\end{table*}

\begin{table*}[ht]
\centering
\small
\begin{tabular}{rrrrrrr}

   & \multicolumn{6}{c}{Manual assessment} \\
  \cmidrule(r{5pt}){2-7}
 Automated assessment & MM & WM & MW & WW & Unknown+M & Unknown+W\\
 \midrule
 MM & 51 & 1 & 1 & 0 & 6 & 0\\
 WM & 0 & 18 & 0 & 1 & 3 & 0\\
 MW & 1 & 0 & 8 & 2 & 0 & 1\\
 WW & 0 & 0 & 1 & 5 & 0 & 1\\
   \bottomrule
\end{tabular}
\captionsetup{width=1\linewidth}
\caption{\textbf{Accuracy of automated gender determination technique for assignment of papers to one of four gender categories in a random sample of 100 papers}.}
\label{tab:tabvalpap}
\end{table*}

\clearpage

\subsubsection*{Imputation of missing data under the null}
The analyses presented in the main text were conducted on the 88\% of papers ($n=54,226$) for which gender could be reliably assigned to both the first and last author. The remaining 12\% of papers, for which the gender of at least one author was unknown, were excluded from the primary analyses. We conducted a secondary analysis in which missing data were imputed and estimates were computed on the full data.

To impute the missing author genders, the gender probabilities obtained from the GAM were used. This model estimated probabilities that a given paper was MM, WM, MW, or WW based on its year of publication, the number of authors, the seniority of the authors, the journal in which it was published, and whether it was a review article. If neither of a paper's authors had an assigned gender, a gender category was randomly assigned to it using the four GAM-estimated probabilities. If one of the paper's authors had an assigned gender, one of the two possible gender categories was randomly assigned to it using the two relevant GAM-estimated probabilities (e.g., if the first author was a woman and the gender of the last author was unknown, either WM or WW would be assigned based on their relative GAM-estimated probabilities). Importantly, this procedure inherently assumes the null of `no effect of author gender on citation patterns' for all imputed data. Thus, the estimates obtained in this sensitivity analysis are likely conservative.

This imputation process, combined with a bootstrapping procedure, was carried out 1000 times. Approximations of the mean effect estimates and their 95\% confidence intervals under this conservative missing data scheme were drawn from the resulting 1000 estimates. The results are shown in Table \ref{tab:tabmiss}, and can be compared to the main paper results found in Table \ref{tab:tabtrue}.

\begin{table*}[ht]
\centering
\scriptsize
\begin{tabular}{rrrrrrrrrrrr}

  & \multicolumn{2}{c}{MM teams} & \multicolumn{2}{c}{W$\cup$W teams} & \multicolumn{2}{c}{WM teams} & \multicolumn{2}{c}{MW teams} & \multicolumn{2}{c}{WW teams} \\\cmidrule(r{5pt}){2-3}\cmidrule(l{5pt}){4-5}\cmidrule(l{5pt}){6-7}\cmidrule(l{5pt}){8-9}\cmidrule(l{5pt}){10-11}
 Effect & Est. & 95\% CI & Est. & 95\% CI & Est. & 95\% CI & Est. & 95\% CI & Est. & 95\% CI \\
 \midrule
MM paper citation rate & +7.7\% & [7.1, 8.3] & +2.8\% & [2.2, 3.4] & +4.1\% & [3.5, 4.8] & +2.9\% & [1.8, 3.9] & -1.5\% & [-2.8, -0.3]\\ 
WM paper citation rate & -9.0\% & [-10.1, -7.8] & -4.8\% & [-5.9, -3.7] & -5.2\% & [-6.6, -3.9] & -6.2\% & [-8.2, -4.2] & -1.9\% & [-4.1, 0.2]\\ 
MW paper citation rate & -7.8\% & [-10.0, -5.8] & +0.0\% & [-2.2, 2.0] & -3.2\% & [-5.8, -0.5] & +1.3\% & [-2.1, 4.7] & +7.4\% & [3.4, 11.6]\\ 
WW paper citation rate & -22.6\% & [-24.8, -20.3] & -5.5\% & [-7.9, -3.0] & -11.5\% & [-14.4, -8.6] & -3.5\% & [-7.9, 1.2] & +9.5\% & [4.0, 15.0]\\ 
MM overcitation trend\\(perc. points per year) & 0.49 & [0.39, 0.60] & 0.27 & [0.16, 0.39] & 0.31 & [0.16, 0.46] & 0.24 & [0.02, 0.45] & 0.26 & [0.01, 0.52]\\ 
Unconditional MM\\overcitation (perc. points) & 5.2 & [4.8, 5.7] & - & - & 3.1 & [2.6, 3.7] & 2.5 & [1.6, 3.3] & -0.4 & [-1.4, 0.5]\\
MM overcitation given\\network (perc. points) & 3.5 & [3.0, 4.0] & - & - & 2.1 & [1.5, 2.7] & 1.9 & [1.0, 2.7] & -0.3 & [-1.2, 0.7]\\
   \bottomrule
\end{tabular}
\captionsetup{width=1\linewidth}
\caption{\textbf{Table of main results after conservative imputation of missing data (no assumed gender imbalance within missing entries)}.}
\label{tab:tabmiss}
\end{table*}

\begin{table*}[ht]
\centering
\scriptsize
\begin{tabular}{rrrrrrrrrrrr}

  & \multicolumn{2}{c}{MM teams} & \multicolumn{2}{c}{W$\cup$W teams} & \multicolumn{2}{c}{WM teams} & \multicolumn{2}{c}{MW teams} & \multicolumn{2}{c}{WW teams} \\\cmidrule(r{5pt}){2-3}\cmidrule(l{5pt}){4-5}\cmidrule(l{5pt}){6-7}\cmidrule(l{5pt}){8-9}\cmidrule(l{5pt}){10-11}
 Effect & Est. & 95\% CI & Est. & 95\% CI & Est. & 95\% CI & Est. & 95\% CI & Est. & 95\% CI \\
 \midrule
MM paper citation rate & +8.0\% & [7.6, 8.5] & +2.5\% & [2.0, 3.1] & +4.0\% & [3.3, 4.7] & +2.7\% & [1.6, 3.7] & -2.0\% & [-3.3, -0.8]\\ 
WM paper citation rate & -9.3\% & [-10.2, -8.3] & -4.6\% & [-5.6, -3.7] & -4.8\% & [-6.0, -3.5] & -6.6\% & [-8.5, -4.6] & -1.8\% & [-4.0, 0.5]\\ 
MW paper citation rate & -9.0\% & [-10.6, -7.4] & -0.1\% & [-2.0, 2.0] & -4.0\% & [-6.3, -1.6] & +2.1\% & [-1.4, 5.7] & +8.2\% & [3.6, 12.5]\\ 
WW paper citation rate & -23.4\% & [-25.4, -21.5] & -4.2\% & [-6.4, -2.0] & -11.4\% & [-14.2, -8.4] & -1.1\% & [-6.2, 3.6] & +12.2\% & [6.7, 17.8]\\ 
MM overcitation trend\\(perc. points per year) & 0.53 & [0.42, 0.63] & 0.29 & [0.17, 0.40] & 0.34 & [0.19, 0.50] & 0.23 & [0.01, 0.48] & 0.25 & [-0.03, 0.53]\\ 
Unconditional MM\\overcitation (perc. points) & 5.5 & [5.1, 5.9] & - & - & 3.1 & [2.6, 3.6] & 2.3 & [1.6, 2.9] & -0.7 & [-1.6, 0.3]\\
MM overcitation given\\network (perc. points) & 3.5 & [3.1, 4.0] & - & - & 2.0 & [1.5, 2.6] & 1.6 & [0.8, 2.3] & -0.3 & [-1.0, 0.6]\\
   \bottomrule
\end{tabular}
\captionsetup{width=1\linewidth}
\caption{\textbf{Table of main results under primary analysis strategy}.}
\label{tab:tabtrue}
\end{table*}

\clearpage

\subsubsection*{Impact of weighting articles by length of reference list}
The primary analyses conducted in this paper are conducted by either passively or actively weighting articles' contributions to estimates by the number of candidate citations present in their reference lists. For the first sections, in which percent over/undercitation is calculated, this weighting is done passively by summing over all citations across citing papers. In the section discussing social network effects, this weighting is done actively by running a quantile regression weighted by the number of citations in an article. Although this weighting was done to stabilize the estimates in the face of articles with very few citations (i.e., percent over/undercitation measures at an individual paper level are highly variable), it was of interest to ensure that the results were robust to this decision. To test for robustness, we re-ran the primary analyses in the paper using unweighted article-level measures of over/undercitation, as opposed to measures that relied on weighting or collapsing across all citations. These results are shown in Table \ref{tab:tabsec}, and can be compared to the main paper results found in Table \ref{tab:tabtrue}.

\begin{table*}[ht]
\centering
\scriptsize
\begin{tabular}{rrrrrrrrrrrr}

  & \multicolumn{2}{c}{MM teams} & \multicolumn{2}{c}{W$\cup$W teams} & \multicolumn{2}{c}{WM teams} & \multicolumn{2}{c}{MW teams} & \multicolumn{2}{c}{WW teams} \\\cmidrule(r{5pt}){2-3}\cmidrule(l{5pt}){4-5}\cmidrule(l{5pt}){6-7}\cmidrule(l{5pt}){8-9}\cmidrule(l{5pt}){10-11}
 Effect & Est. & 95\% CI & Est. & 95\% CI & Est. & 95\% CI & Est. & 95\% CI & Est. & 95\% CI \\
 \midrule
MM paper citation rate & +7.1\% & [6.4, 7.7] & +1.5\% & [0.8, 2.2] & +2.9\% & [2.0, 3.8] & +2.2\% & [0.8, 3.6] & -3.0\% & [-4.6, -1.4]\\ 
WM paper citation rate & -8.8\% & [-10.0, -7.6] & -3.2\% & [-4.5, -1.9] & -3.5\% & [-5.2, -1.8] & -6.1\% & [-8.6, -3.5] & +1.0\% & [-2.1, 4.0]\\ 
MW paper citation rate & -6.8\% & [-9.0, -4.5] & +0.2\% & [-1.9, 2.6] & -2.8\% & [-5.9, 0.5] & +4.2\% & [-0.9, 9.1] & +4.3\% & [-1.0, 9.5]\\ 
WW paper citation rate & -21.9\% & [-24.5, -19.3] & -4.7\% & [-7.4, -2.0] & -11.7\% & [-15.1, -8.1] & -0.4\% & [-6.5, 5.4] & +10.4\% & [4.2, 16.7]\\ 
MM overcitation trend\\(perc. points per year) & 0.39 & [0.25, 0.52] & 0.21 & [0.07, 0.35] & 0.17 & [-0.03, 0.35] & 0.26 & [-0.04, 0.56] & 0.25 & [-0.08, 0.58]\\ 
Unconditional MM\\overcitation (perc. points) & 5.1 & [4.7, 5.5] & - & - & 2.5 & [1.9, 3.0] & 2.2 & [1.4, 2.9] & -0.8 & [-1.7, 0.0]\\
MM overcitation given\\network (perc. points) & 3.3 & [2.8, 3.7] & - & - & 1.6 & [1.0, 2.2] & 1.8 & [1.0, 2.5] & -0.4 & [-1.3, 0.5]\\
   \bottomrule
\end{tabular}
\captionsetup{width=1\linewidth}
\caption{\textbf{Table of main results using unweighted article-level assessments of over/under citation}.}
\label{tab:tabsec}
\end{table*}

\clearpage

\subsubsection*{Rationale and impact of self-citation exclusion}
\begin{figure*}[ht]
\centering
\includegraphics[width=.65\linewidth]{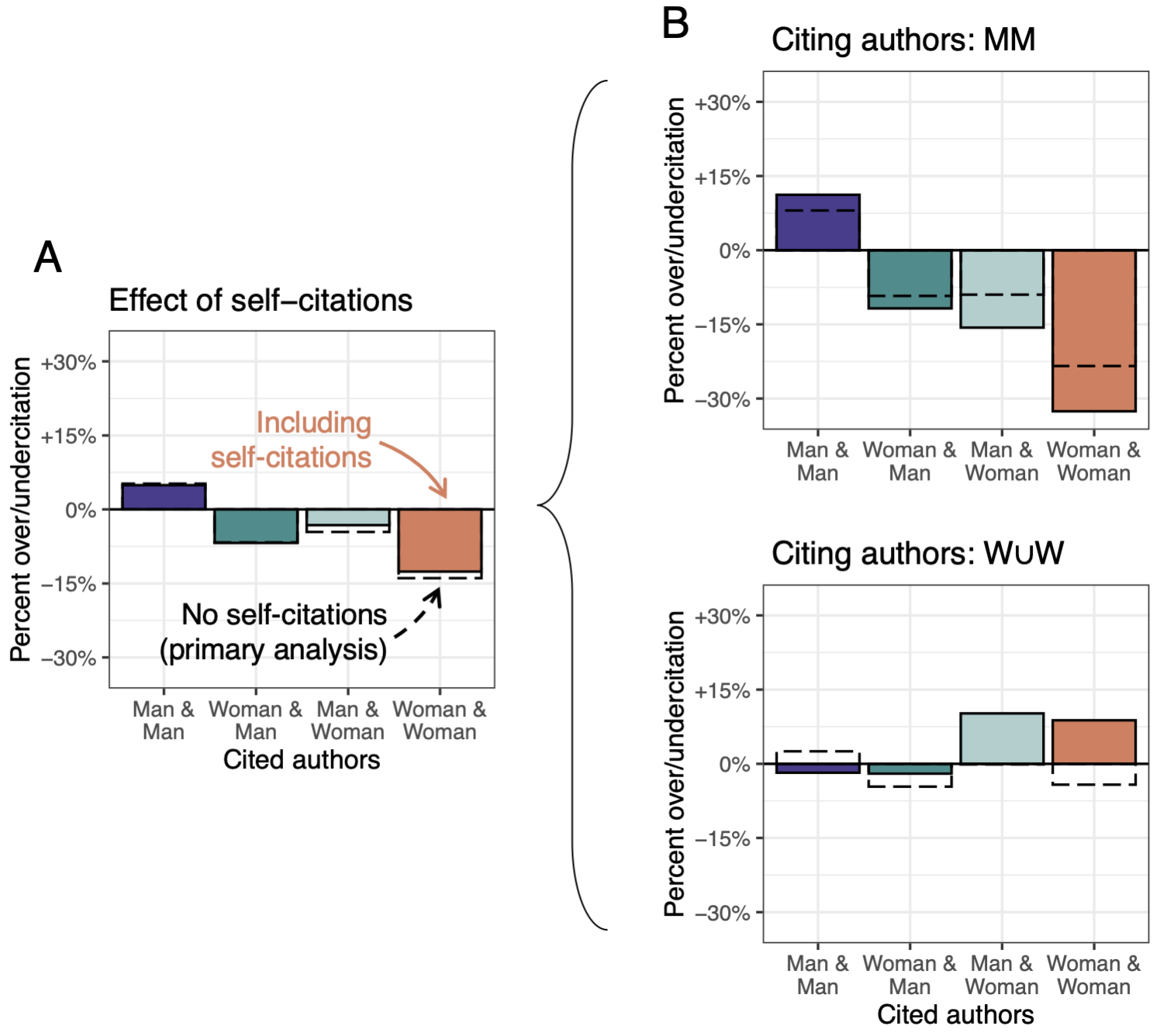}
\caption{\textbf{Effect of self-citations on the distribution of citations across gender categories.} \textit{(A)} Overall over/undercitation of author gender categories excluding self-citations (dotted) and including self-citations (colored). \textit{(B)} Over/undercitation of author gender categories excluding and including self-citations within MM reference lists (top) and W$\cup$W reference lists (bottom).}
\label{fig:FigureS1}
\end{figure*}

For all analyses reported in the primary manuscript, self-citations were removed from consideration. Although the potential for gendered patterns in self-citations is potentially interesting, their removal in this study allowed us to isolate more comparable aspects of citation behavior between men and women in the field. Here we show additional results related to self-citation behavior. First, to assess the impact that self-citations have on gendered citation behavior, we conducted the primary analyses of over/undercitation without removing instances of self-citation. Here, we find that the inclusion of self-citations does not make a meaningful difference in the overall over/undercitation patterns. Their inclusion does, however, create even larger differences between the citation behavior of MM and W$\cup$W teams (Figure \ref{fig:FigureS1}), as self-citations necessarily increase the citation rate of men-led papers within MM reference lists, and increase the citation rate of women-led papers within W$\cup$W reference lists. To ensure that missed self-citations were not driving the observed differences, we removed all citations in which the cited first or last author had the same last name as either the first or last author of the citing paper (i.e., a very conservative removal of potential self-citations). This more expansive removal led to negligible change in the primary results, suggesting that missed instances of self-citation were rare (Table \ref{tab:tabdisam}).

Additionally, we examined potential gendered differences in the rate of self-citations across author genders. While a more thorough version of this analysis was carried out in King et al., 2017 ("Men Set Their Own Cites High"), it may be of interest to the field to see the results for neuroscience. We found that as a proportion of reference list length, MM and WM teams tended to self-cite at higher rates than MW and WW teams. However, as a proportion of potential self-citations (i.e., count of authors' previous citable papers), rates of self-citation were relatively similar across groups (though still slightly higher in MM than WW; Table \ref{tab:tabself}).

\begin{table*}[ht]
\centering
\footnotesize
\begin{tabular}{rrrrrrr}

  & \multicolumn{2}{c}{Overall} & \multicolumn{2}{c}{MM teams} & \multicolumn{2}{c}{W$\cup$W teams} \\\cmidrule(r{5pt}){2-3}\cmidrule(l{5pt}){4-5}\cmidrule(l{5pt}){6-7}
 Effect & Primary & Strict & Primary & Strict & Primary & Strict \\
 \midrule
MM paper citation rate & +5.21\% & +5.21\%  & +8.03\% & +8.01\% & +2.53\% & +2.56\%\\ 
WM paper citation rate & -6.69\% & -6.69\% & -9.26\% & -9.22\% & -4.63\% & -4.65\%\\ 
MW paper citation rate & -4.57\% & -4.59\% & -9.00\% & -8.96\% & -0.12\% & -0.16\%\\ 
WW paper citation rate & -13.92\% & -13.95\% & -23.43\% & -23.39\% & -4.23\% & -4.36\%\\ 
   \bottomrule
\end{tabular}
\captionsetup{width=1\linewidth}
\caption{\textbf{Potential impact of missed instances of self-citation}.}
\label{tab:tabdisam}
\end{table*}

\begin{table*}[ht]
\centering
\scriptsize
\begin{tabular}{rrrrrrrrrr}

  & \multicolumn{2}{c}{MM teams} & \multicolumn{2}{c}{WM teams} & \multicolumn{2}{c}{MW teams} & \multicolumn{2}{c}{WW teams} \\\cmidrule(r{5pt}){2-3}\cmidrule(l{5pt}){4-5}\cmidrule(l{5pt}){6-7}\cmidrule(l{5pt}){8-9}
 Effect & Est. & 95\% CI & Est. & 95\% CI & Est. & 95\% CI & Est. & 95\% CI \\
 \midrule
Mean self-citations by\\reference list length & 13.5\% & [13.2, 13.7] & 13.6\% & [13.2, 14.0] & 12.1\% & [11.5, 12.6] & 10.8\% & [10.2, 11.3]\\
Median self-citations by\\reference list length & 9.1\% & [8.7, 9.1] & 8.5\% & [8.3, 9.1] & 7.2\% & [6.7, 7.7] & 6.3\% & [5.6, 7.1]\\
Mean self-citations by\\num. previous papers & 17.0\% & [16.7, 17.3] & 17.0\% & [16.6, 17.4] & 17.8\% & [17.2, 18.5] & 17.1\% & [16.3, 17.8]\\
Median self-citations by\\num. previous papers & 12.5\% & [11.8, 12.5] & 11.7\% & [11.1, 12.5] & 13.0\% & [12.1, 14.3] & 10.9\% & [9.1, 12.5]\\
   \bottomrule
\end{tabular}
\captionsetup{width=1\linewidth}
\caption{\textbf{Comparison of self-citation rates across author gender categories, relative to both the length of reference lists and the number of potential self-citations}.}
\label{tab:tabself}
\end{table*}

\clearpage

\subsubsection*{Effect of varying definitions of self-citation}
For the primary analyses, we defined self-citations as papers for which either the cited first or last author was the first or last author on the citing paper (according to the previously described author name disambiguation procedure). Although this is a somewhat restrictive definition, it is the only type of self-citation for which the author gender of the cited paper is necessarily determined by the author gender of the citing paper. Here, we determine whether using broader definitions of self-citation has a meaningful effect on the results. We consider two alternate definitions of self-citation. \textit{Broad - Citing} defines self-citations as papers for which either the cited first or last author was a co-author (not exclusively first/last) on the citing paper. \textit{Broad - Cited} defines self-citations as cited papers for which any co-authors (not exclusively first/last) were first or last author on the citing paper. Tables \ref{tab:tabentlist} and \ref{tab:tabcoauth} illustrate that all possible definitions yielded highly similar results.

\begin{table*}[ht]
\centering
\footnotesize
\begin{tabular}{rrrrrrr}

  & \multicolumn{2}{c}{Overall} & \multicolumn{2}{c}{MM teams} & \multicolumn{2}{c}{W$\cup$W teams} \\\cmidrule(r{5pt}){2-3}\cmidrule(l{5pt}){4-5}\cmidrule(l{5pt}){6-7}
 Effect & Primary & Broad - Citing & Primary & Broad - Citing & Primary & Broad - Citing \\
 \midrule
MM paper citation rate & +5.21\% & +5.18\%  & +8.03\% & +8.01\% & +2.53\% & +2.48\%\\ 
WM paper citation rate & -6.69\% & -6.83\% & -9.26\% & -9.32\% & -4.63\% & -4.87\%\\ 
MW paper citation rate & -4.57\% & -4.26\% & -9.00\% & -8.89\% & -0.12\% & +0.52\%\\ 
WW paper citation rate & -13.92\% & -13.71\% & -23.43\% & -23.35\% & -4.23\% & -3.90\%\\ 
   \bottomrule
\end{tabular}
\captionsetup{width=1\linewidth}
\caption{\textbf{Effect of expanding the definition of self-citation to include additional papers}. Primary analysis defined self-citation as cited papers for which the first or last author is the first or last author on the citing paper. ``Broad - Citing" analysis here refers to the process of defining self-citations as any papers for which the \textit{cited} first or last author is a co-author on the \textit{citing} paper.}
\label{tab:tabentlist}
\end{table*}

\begin{table*}[ht]
\centering
\footnotesize
\begin{tabular}{rrrrrrr}

  & \multicolumn{2}{c}{Overall} & \multicolumn{2}{c}{MM teams} & \multicolumn{2}{c}{W$\cup$W teams} \\\cmidrule(r{5pt}){2-3}\cmidrule(l{5pt}){4-5}\cmidrule(l{5pt}){6-7}
 Effect & Primary & Broad - Cited & Primary & Broad - Cited & Primary & Broad - Cited \\
 \midrule
MM paper citation rate & +5.21\% & +5.25\%  & +8.03\% & +8.05\% & +2.53\% & +2.64\%\\ 
WM paper citation rate & -6.69\% & -6.83\% & -9.26\% & -9.27\% & -4.63\% & -5.00\%\\ 
MW paper citation rate & -4.57\% & -4.49\% & -9.00\% & -9.11\% & -0.12\% & +0.27\%\\ 
WW paper citation rate & -13.92\% & -14.14\% & -23.43\% & -23.77\% & -4.23\% & -4.42\%\\ 
   \bottomrule
\end{tabular}
\captionsetup{width=1\linewidth}
\caption{\textbf{Effect of expanding the definition of self-citation to include additional papers}. Primary analysis defined self-citation as cited papers for which the first or last author is the first or last author on the citing paper. ``Broad - Cited" analysis here refers to the process of defining self-citations as any papers for which the \textit{citing} first or last author is a co-author on the \textit{cited} paper.}
\label{tab:tabcoauth}
\end{table*}

\clearpage

\subsubsection*{Construction of graph-preserving null model}
\begin{figure*}[ht]
\centering
\includegraphics[width=\linewidth]{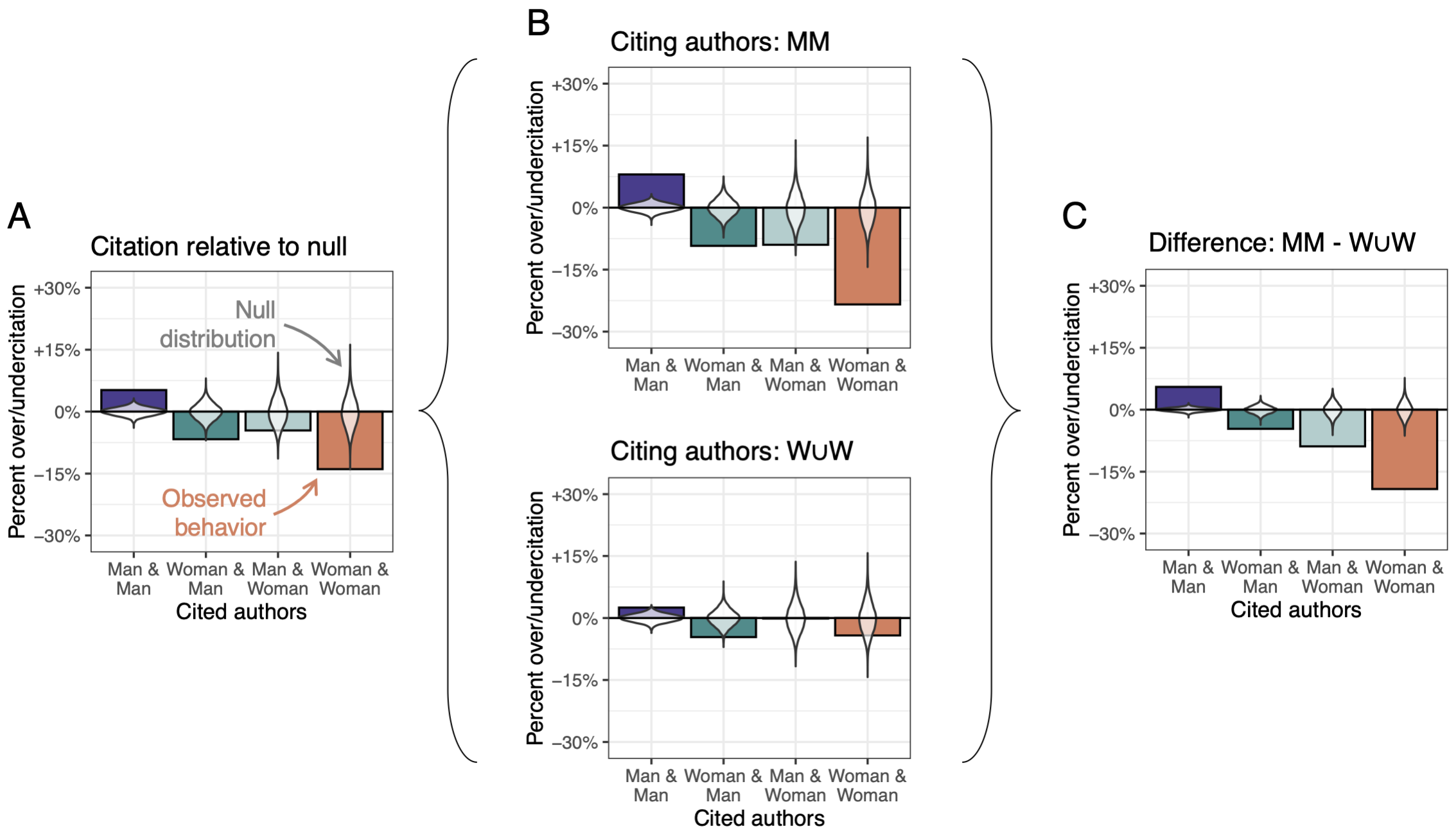}
\caption{\textbf{Visualization of the distribution of citation behavior under the null hypothesis that citation is independent of cited author gender conditional on specific paper characteristics.} \textit{(A)} Overall over/undercitation of author gender categories (colored bars) with the graph-preserving null distributions overlaid (translucent violins). \textit{(B)} Over/undercitation of author gender categories and overlaid null distributions within MM reference lists (top) and W$\cup$W reference lists (bottom). \textit{(C)} Difference between citation rates within MM and W$\cup$W reference lists and overlaid null distributions.}
\label{fig:FigureS2}
\end{figure*}

To determine the extent to which observed effects could be explained by the structure of the citation graph, a graph null model was constructed to assess significance of primary analyses. This null model was derived from the randomization of \textit{cited} papers' author gender categories. To incorporate other characteristics of papers that may be relevant to citation behavior, randomization was carried out by probabilistically drawing new gender categories for each paper according to their GAM-estimated gender probabilities. Randomly sampling gender categories for each paper therefore produces a null model in which cited author gender is conditionally independent of citation rates and citing author gender (conditional on the characteristics included in the GAM model), while the structure of the citation graph and the long-tailed nature of the citation distribution are both preserved. 10,000 randomizations were carried out, and \textit{p}-values were calculated by taking the proportion of randomizations for which the absolute value of a given null estimate was greater than the absolute value of its respective observed estimate. To account for multiple comparisons, reported \textit{p}-values were corrected according to the Holm-Bonferroni procedure. Figure \ref{fig:FigureS2} shows the null distributions for the primary citation imbalance measures overlaid on the observed values. Table \ref{tab:tabnull} for means and standard errors of all primary analyses across the 10,000 randomizations).

\vspace{5cm}
\begin{table*}[ht]
\centering
\normalsize
\begin{tabular}{rrrrr}

 Effect & Est. & Null mean & Null SE & Null 2.5/97.5\%-interval  \\
 \midrule
Overall - MM citation rate & 5.21 & 0.00 & 0.93 & [-1.84, 1.80]\\
Overall - WM citation rate & -6.69 & 0.00 & 1.93 & [-3.67, 3.85]\\
Overall - MW citation rate & -4.57 & 0.04 & 3.38 & [-6.25, 7.05]\\
Overall - WW citation rate & -13.92 & -0.05 & 3.73 & [-7.06, 7.65]\\
MM ref. list - MM citation rate & 8.03 & 0.00 & 0.97 & [-1.89, 1.85]\\
MM ref. list - WM citation rate & -9.26 & 0.00 & 2.01 & [-3.80, 4.02]\\
MM ref. list - MW citation rate & -9.00 & 0.04 & 3.55 & [-6.63, 7.39]\\
MM ref. list - WW citation rate & -23.43 & -0.04 & 3.96 & [-7.50, 8.05]\\
W$\cup$W ref. list - MM citation rate & 2.53 & 0.00 & 0.95 & [-1.88, 1.85]\\
W$\cup$W ref. list - WM citation rate & -4.63 & 0.01 & 1.93& [-3.66, 3.83]\\
W$\cup$W ref. list - MW citation rate & -0.12 & 0.04 & 3.37 & [-6.24, 7.07]\\
W$\cup$W ref. list - WW citation rate & -4.23 & -0.06 & 3.73 & [-7.06, 7.53]\\
MM/W$\cup$W diff - MM citation rate & 5.50 & 0.00 & 0.43 & [-0.83, 0.83]\\
MM/W$\cup$W diff - WM citation rate & -4.63 & -0.01 & 0.87 & [-1.71, 1.69]\\
MM/W$\cup$W diff - MW citation rate & -8.88 & 0.00 & 1.58 & [-3.13, 3.07]\\
MM/W$\cup$W diff - WW citation rate & -19.19 & 0.02 & 1.86 & [-3.56, 3.71]\\
Overall - MM citation trend & 0.41 & 0.00 & 0.09 & [-0.17, 0.17]\\
MM ref. list - MM citation trend & 0.54 & 0.00 & 0.10 & [-0.19, 0.19]\\
W$\cup$W ref. list - MM citation trend & 0.29 & 0.00 & 0.09 & [-0.17, 0.17]\\
MM/W$\cup$W diff - MM citation trend & 0.25 & 0.00 & 0.07 & [-0.15, 0.15]\\
MM - MM uncond. overcitation & 5.46 & 0.18 & 0.61 & [-1.04, 1.38]\\
WM - MM uncond. overcitation & 3.04 & 0.19 & 0.64 & [-1.08, 1.45]\\
MW - MM uncond. overcitation & 2.39 & 0.21 & 0.70 & [-1.14, 1.56]\\
WW - MM uncond. overcitation & -0.71 & 0.11 & 0.68 & [-1.15, 1.50]\\
MM - MM network overcitation & 3.50 & 0.19 & 0.61 & [-1.02, 1.39]\\
WM - MM network overcitation & 1.92 & 0.21 & 0.64 & [-1.06, 1.43]\\
MW - MM network overcitation & 1.60 & 0.22 & 0.71 & [-1.18, 1.58]\\
WW - MM network overcitation & -0.38 & 0.12 & 0.69 & [-1.16, 1.53]\\
   \bottomrule
\end{tabular}
\captionsetup{width=1\linewidth}
\caption{\textbf{Means, standard errors, and 2.5/97.5-percentiles of primary estimates over 10,000 randomizations of the graph-preserving null model}.}
\label{tab:tabnull}
\end{table*}

\clearpage

\subsubsection*{Potential confounding impact of research subfields}
\begin{figure*}[ht]
\centering
\includegraphics[width=.65\linewidth]{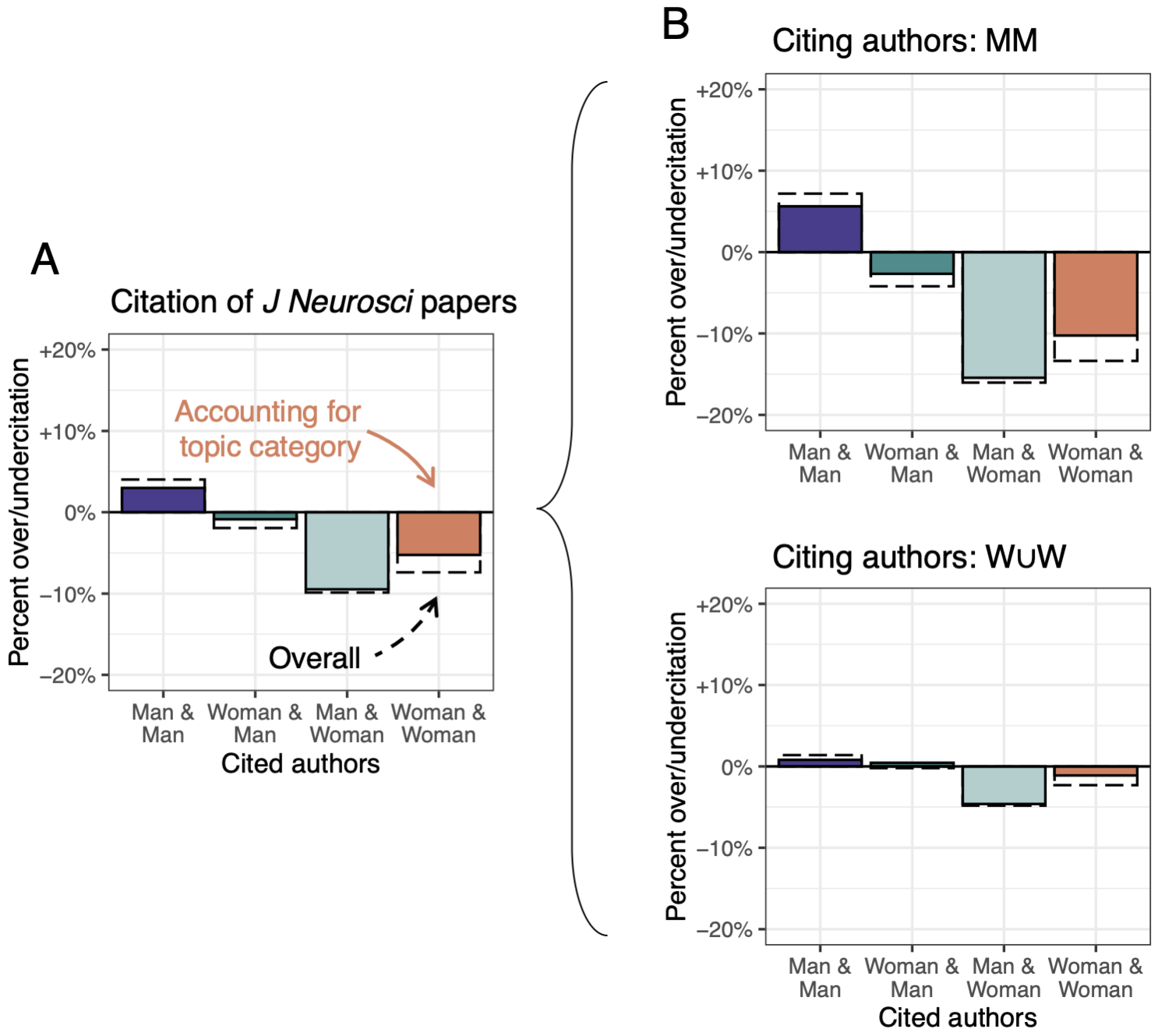}
\caption{\textbf{Impact of research subfields on citation behavior, using a subset of \textit{Journal of Neuroscience} papers with sub-disciplinary classifications.} \textit{(A)} Overall over/undercitation of author gender categories before accounting for subfield (dotted) and after accounting for subfield (colored). \textit{(B)} Over/undercitation of author gender categories before and after accounting for subfield within MM reference lists (top) and W$\cup$W reference lists (bottom).}
\label{fig:FigureS3}
\end{figure*}

Though the role of neuroscience subfields is not directly accounted for within the primary analyses, it is important to understand whether and to what extent relationships between gender and subfield confound our results. To assess this possibility, we conducted a sensitivity analysis on a subset of \textit{Journal of Neuroscience} papers. \textit{Journal of Neuroscience} was chosen because it contains the most articles within our dataset, and because it uses a consistent sub-disciplinary classification scheme for its papers. Specifically, almost all of its papers are classified as either \textit{behavioral/systems}, \textit{systems/circuits}, \textit{neurobiology of disease}, \textit{development/plasticity/repair}, \textit{behavioral/systems/cognitive}, or \textit{cellular/molecular}. Two separate author gender-predicting generalized additive models (GAMs) were fit to the subset of 21,338 articles with one of these classifications. The first did not include the subfield classification, and the second did. Estimates of the over/undercitation of author genders within these 21,338 articles were then calculated as per the primary analyses using each estimated model. Figure \ref{fig:FigureS3} shows the results, with dashed lines indicating the results without subfield classifications (i.e., the same model as was used in the primary manuscript), and colored bars indicating results after accounting for subfields. The results suggest that subfields likely have little impact on either the extent of citation imbalance or the discrepancy in citation behavior across citing author genders.

\clearpage

\subsubsection*{Potential differential effects among high- and low-citation papers}
\begin{figure*}[ht]
\centering
\includegraphics[width=.65\linewidth]{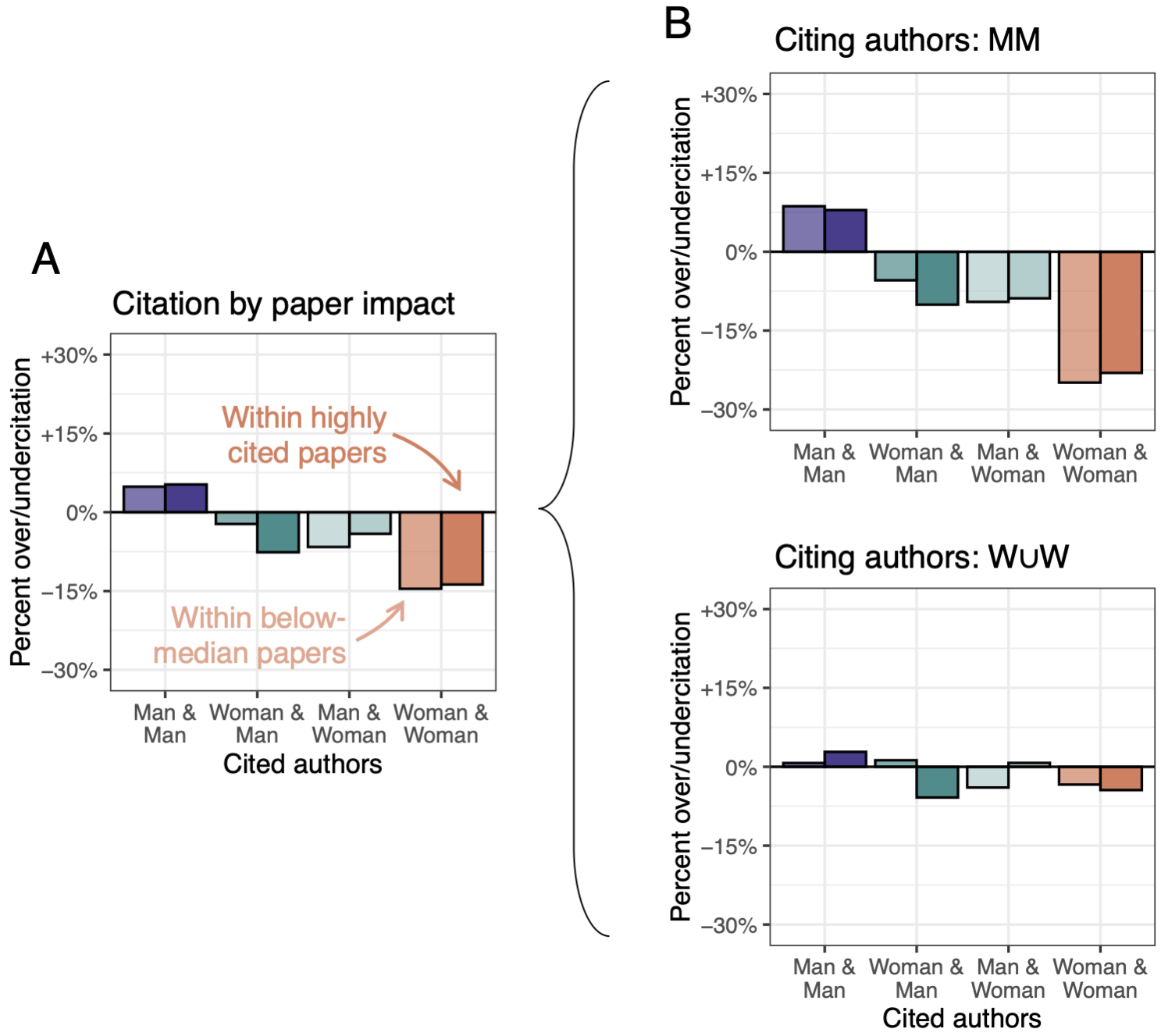}
\caption{\textbf{Citation behavior after stratifying cited papers by their status at the top or bottom of the citation distribution (by median split).} \textit{(A)} Overall over/undercitation of author gender categories among low-citation (left, transparent) papers and high-citation (right, opaque) papers. \textit{(B)} Over/undercitation of author gender categories among low-citation papers and high-citation papers within MM reference lists (top) and W$\cup$W reference lists (bottom).}
\label{fig:FigureS4}
\end{figure*}

Because the citation distribution is long-tailed, it is of particular interest to determine whether the observed overcitation of men is driven primarily by a subset of highly cited papers, over whether over/undercitation occurs relatively consistently for high- and low-citation papers. To address this question, we separately quantified over/undercitation of author gender categories within the top 50\% and bottom 50\% of the citation distribution. In other words, papers in the data were separated into two groups according to a median split (the median number of citations at time of data retrieval was 50), and citation rates relative to expectation were assessed within each group according to the same procedure used for the primary analyses. While the papers in the top half of the distribution accounted for three-quarters of all citations in the data, the patterns of over/undercitation were similar for both sets (Figure \ref{fig:FigureS4}). Specifically, below-median MM/WM/MW/WW papers were cited +4.8\%, -2.2\%, -6.6\%, and -14.6\% relative to expectation, while above-median papers were cited +5.3\%, -7.6\%, -4.1\%, and -13.8\% relative to expectation. This suggests that the observed imbalance is not driven by a `rich club' of highly cited papers by men.

\clearpage

\subsubsection*{Role of seniority/productivity in network structure and MM overcitation}
Scholars' seniority and productivity have a potentially important role in constructing the collaboration network and shaping citation behavior. Therefore, it was of interest to determine the extent to which the most senior (i.e., in our data, most highly productive) scientists drive the network homophily and citation patterns. To assess this possibility, we performed a median split of articles based on the total number of papers the first/last-author team had published within the context of our dataset (i.e., top 5 journals, 1995-2018). For the purpose of the analysis, teams who published $>8$ papers in this timespan are referred to as \textit{more productive} and teams who published $\leq8$ papers are referred to as \textit{less productive}.

\begin{figure*}[ht]
\centering
\includegraphics[width=.65\linewidth]{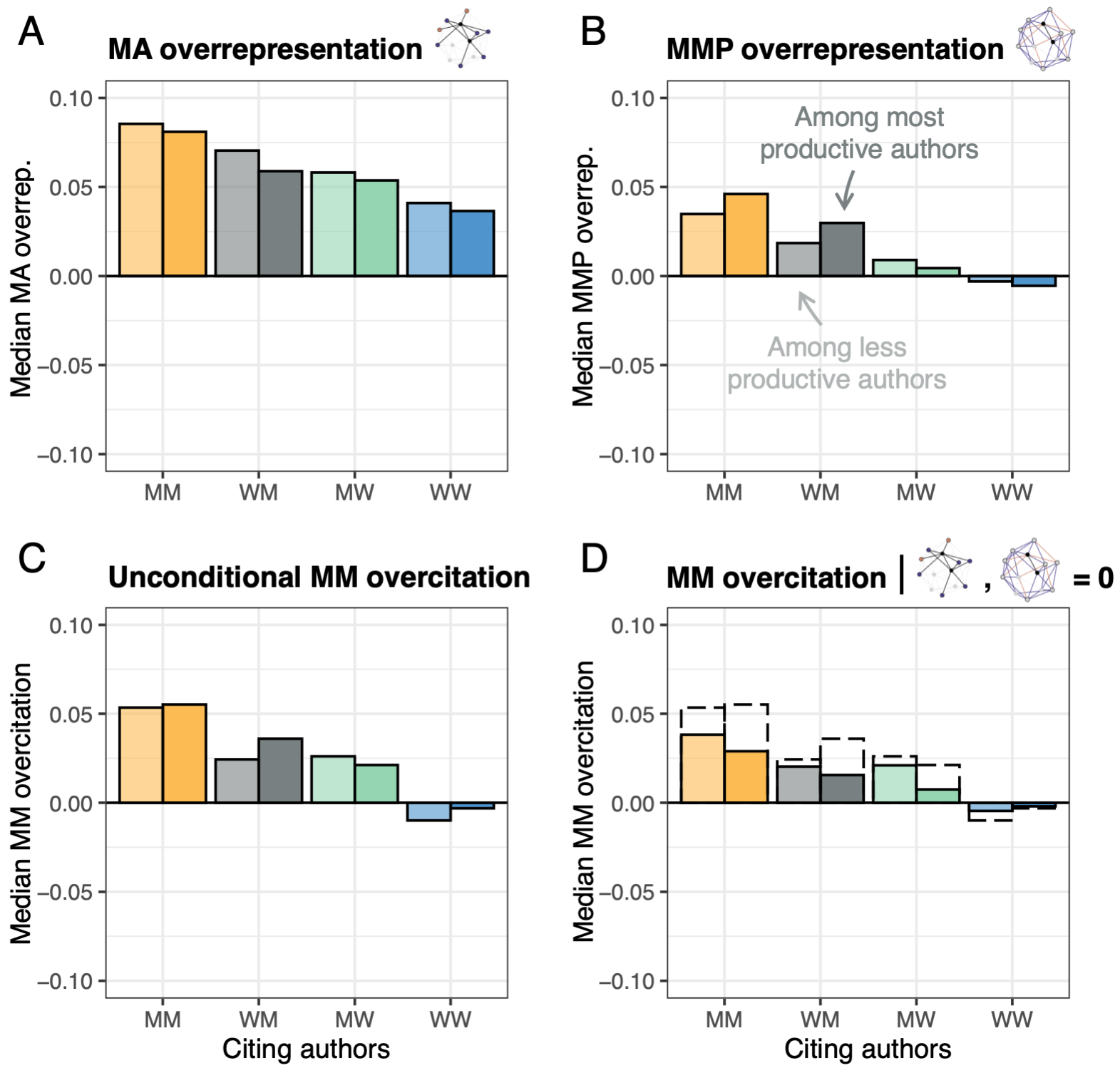}
\caption{\textbf{Relationship between seniority/productivity, network structure, and citation patterns.} \textit{(A)} Man-author overrepresentation by author gender among low productivity (left, transparent) and high productivity (right, opaque) teams. \textit{(B)} MM-paper overrepresentation by author gender among low productivity and high productivity teams. \textit{(C)} Overcitation of MM papers (not accounting for network structure) by author gender among low productivity and high productivity teams. Shows similar patterns of citation between low and high seniority/productivity teams. \textit{(D)} Overcitation of MM papers by author gender, after accounting for network structure. Shows that network structure explains more of the citation imbalance within high productivity teams than low productivity teams.}
\label{fig:FigureS5}
\end{figure*}

The network structure showed subtle differences between the two groups. More productive teams tended to have very slightly less man-author overrepresentation within their networks (Figure \ref{fig:FigureS5}A). Additionally, more productive teams tended to have higher MM-paper overrepresentation in their networks when men were the last-author, and lower MM-paper overrepresentation when women were last-author (Figure \ref{fig:FigureS5}B). In terms of citation patterns, higher and lower productivity teams tended to show similar citation behavior across author genders both before (Figure \ref{fig:FigureS5}C) and after accounting for network effects (Figure \ref{fig:FigureS5}D). Yet interestingly, the network structure of more productive teams appeared to be more strongly associated with citation behavior than that of less productive teams (Figure \ref{fig:FigureS5}D). Specifically, across all papers, a one percentage point increase in MA overrepresentation corresponded to a 0.09 percentage point increase in MM overcitation. When broken down by less/more productive teams, this value was 0.05 and 0.15, respectively. Similarly, across all papers, a one percentage point increase in MM-paper overrepresentation corresponded to a 0.24 percentage point increase in MM overcitation. When broken down by less/more productive teams, this value was 0.21 and 0.28, respectively. This finding may suggest that more senior authors’ citation behavior is more closely tied to the structure of their social networks.

\clearpage

\subsubsection*{Quantile regression vs. linear regression for estimating MM overcitation}
The analyses in the ``The relationship between social networks and citation behavior" section were conducted using quantile regression to fit the conditional median of the outcome. Quantile regression was chosen, as opposed to linear regression, because of the bounded and skewed nature of the outcome measure. However, to ensure that the results are robust to this modeling choice, here we compare the results of that section using quantile regression to the results that would be obtained using linear regression. Table \ref{tab:tablin} shows that the results, while somewhat attenuated using linear regression, are consistent with the results obtained with quantile regression.

\begin{table*}[ht]
\centering
\footnotesize
\begin{tabular}{rrrrrrrrrr}

  & \multicolumn{2}{c}{MM teams} & \multicolumn{2}{c}{WM teams} & \multicolumn{2}{c}{MW teams} & \multicolumn{2}{c}{WW teams} \\\cmidrule(r{5pt}){2-3}\cmidrule(l{5pt}){4-5}\cmidrule(l{5pt}){6-7}\cmidrule(l{5pt}){8-9}
 Effect & Est. & 95\% CI & Est. & 95\% CI & Est. & 95\% CI & Est. & 95\% CI \\
 \midrule
Unconditional \textit{median} MM\\overcitation (perc. points) & 5.5 & [5.1, 5.9] & 3.1 & [2.6, 3.6] & 2.3 & [1.6, 3.0] & -0.7 & [-1.6, 0.3]\\
\textit{Median} MM overcitation given\\network (perc. points) & 3.5 & [3.1, 3.9] & 2.0 & [1.5, 2.6] & 1.6 & [0.8, 2.3] & -0.3 & [-1.0, 0.6]\\
Unconditional \textit{mean} MM\\overcitation (perc. points) & 4.7 & [4.4, 5.0] & 2.4 & [2.0, 2.8] & 1.6 & [0.9, 2.2] & -1.2 & [-2.0, -0.4]\\
\textit{Mean} MM overcitation given\\network (perc. points) & 3.1 & [2.8, 3.4] & 1.3 & [0.9, 1.8] & 1.2 & [0.6, 1.9] & -1.2 & [-2.0, -0.4]\\
   \bottomrule
\end{tabular}
\captionsetup{width=1\linewidth}
\caption{\textbf{Comparison between overcitation results obtained using quantile regression and linear regression}.}
\label{tab:tablin}
\end{table*}



\section*{References}
\begin{thebibliography}{73}
\providecommand{\natexlab}[1]{#1}
\providecommand{\url}[1]{\texttt{#1}}
\expandafter\ifx\csname urlstyle\endcsname\relax
  \providecommand{\doi}[1]{doi: #1}\else
  \providecommand{\doi}{doi: \begingroup \urlstyle{rm}\Url}\fi

\bibitem[Holman et~al.(2018)Holman, Stuart-Fox, and Hauser]{holman_gender_2018}
Luke Holman, Devi Stuart-Fox, and Cindy~E. Hauser.
\newblock The gender gap in science: {How} long until women are equally
  represented?
\newblock \emph{PLOS Biology}, 16\penalty0 (4):\penalty0 e2004956, April 2018.
\newblock ISSN 1545-7885.
\newblock \doi{10.1371/journal.pbio.2004956}.

\bibitem[Moss-Racusin et~al.(2012)Moss-Racusin, Dovidio, Brescoll, Graham, and
  Handelsman]{moss-racusin_science_2012}
Corinne~A. Moss-Racusin, John~F. Dovidio, Victoria~L. Brescoll, Mark~J. Graham,
  and Jo~Handelsman.
\newblock Science faculty's subtle gender biases favor male students.
\newblock \emph{Proceedings of the National Academy of Sciences}, 109\penalty0
  (41):\penalty0 16474--16479, October 2012.
\newblock ISSN 0027-8424, 1091-6490.
\newblock \doi{10.1073/pnas.1211286109}.

\bibitem[Bornmann et~al.(2007)Bornmann, Mutz, and Daniel]{bornmann_gender_2007}
Lutz Bornmann, Rudiger Mutz, and Hans-Dieter Daniel.
\newblock Gender differences in grant peer review: {A} meta-analysis.
\newblock \emph{Journal of Informetrics}, 1\penalty0 (3):\penalty0 226--238,
  July 2007.
\newblock ISSN 17511577.
\newblock \doi{10.1016/j.joi.2007.03.001}.

\bibitem[Jagsi(2009)]{jagsi_sex_2009}
Reshma Jagsi.
\newblock Sex {Differences} in {Attainment} of {Independent} {Funding} by
  {Career} {Development} {Awardees}.
\newblock \emph{Annals of Internal Medicine}, 151\penalty0 (11):\penalty0 804,
  December 2009.
\newblock ISSN 0003-4819.
\newblock \doi{10.7326/0003-4819-151-11-200912010-00009}.

\bibitem[van~der Lee and Ellemers(2015)]{van_der_lee_gender_2015}
Romy van~der Lee and Naomi Ellemers.
\newblock Gender contributes to personal research funding success in {The}
  {Netherlands}.
\newblock \emph{Proceedings of the National Academy of Sciences}, 112\penalty0
  (40):\penalty0 12349--12353, October 2015.
\newblock ISSN 0027-8424, 1091-6490.
\newblock \doi{10.1073/pnas.1510159112}.

\bibitem[Sarsons(2017)]{sarsons_recognition_2017}
Heather Sarsons.
\newblock Recognition for {Group} {Work}: {Gender} {Differences} in {Academia}.
\newblock \emph{American Economic Review}, 107\penalty0 (5):\penalty0 141--145,
  May 2017.
\newblock ISSN 0002-8282.
\newblock \doi{10.1257/aer.p20171126}.

\bibitem[MacNell et~al.(2015)MacNell, Driscoll, and Hunt]{macnell_whats_2015}
Lillian MacNell, Adam Driscoll, and Andrea~N. Hunt.
\newblock What’s in a {Name}: {Exposing} {Gender} {Bias} in {Student}
  {Ratings} of {Teaching}.
\newblock \emph{Innovative Higher Education}, 40\penalty0 (4):\penalty0
  291--303, August 2015.
\newblock ISSN 0742-5627, 1573-1758.
\newblock \doi{10.1007/s10755-014-9313-4}.

\bibitem[Mengel et~al.(2019)Mengel, Sauermann, and Zölitz]{mengel_gender_2019}
Friederike Mengel, Jan Sauermann, and Ulf Zölitz.
\newblock Gender {Bias} in {Teaching} {Evaluations}.
\newblock \emph{Journal of the European Economic Association}, 17\penalty0
  (2):\penalty0 535--566, April 2019.
\newblock ISSN 1542-4766, 1542-4774.
\newblock \doi{10.1093/jeea/jvx057}.

\bibitem[Boring(2017)]{boring_gender_2017}
Anne Boring.
\newblock Gender biases in student evaluations of teaching.
\newblock \emph{Journal of Public Economics}, 145:\penalty0 27--41, January
  2017.
\newblock ISSN 00472727.
\newblock \doi{10.1016/j.jpubeco.2016.11.006}.

\bibitem[Nielsen(2016)]{nielsen_limits_2016}
Mathias~W. Nielsen.
\newblock Limits to meritocracy? {Gender} in academic recruitment and promotion
  processes.
\newblock \emph{Science and Public Policy}, 43\penalty0 (3):\penalty0 386--399,
  June 2016.
\newblock ISSN 0302-3427, 1471-5430.
\newblock \doi{10.1093/scipol/scv052}.

\bibitem[Van~den Brink(2011)]{van_den_brink_scouting_2011}
Marieke Van~den Brink.
\newblock Scouting for talent: {Appointment} practices of women professors in
  academic medicine.
\newblock \emph{Social Science \& Medicine}, 72\penalty0 (12):\penalty0
  2033--2040, June 2011.
\newblock ISSN 02779536.
\newblock \doi{10.1016/j.socscimed.2011.04.016}.

\bibitem[De~Paola and Scoppa(2015)]{de_paola_gender_2015}
Maria De~Paola and Vincenzo Scoppa.
\newblock Gender {Discrimination} and {Evaluators}’ {Gender}: {Evidence} from
  {Italian} {Academia}.
\newblock \emph{Economica}, 82\penalty0 (325):\penalty0 162--188, January 2015.
\newblock ISSN 00130427.
\newblock \doi{10.1111/ecca.12107}.

\bibitem[West et~al.(2013)West, Jacquet, King, Correll, and
  Bergstrom]{west_role_2013}
Jevin~D. West, Jennifer Jacquet, Molly~M. King, Shelley~J. Correll, and Carl~T.
  Bergstrom.
\newblock The {Role} of {Gender} in {Scholarly} {Authorship}.
\newblock \emph{PLoS ONE}, 8\penalty0 (7):\penalty0 e66212, July 2013.
\newblock ISSN 1932-6203.
\newblock \doi{10.1371/journal.pone.0066212}.

\bibitem[Wilhelm et~al.(2018)Wilhelm, Conklin, and Hassoun]{wilhelm_new_2018}
Isaac Wilhelm, Sherri~Lynn Conklin, and Nicole Hassoun.
\newblock New data on the representation of women in philosophy journals:
  2004–2015.
\newblock \emph{Philosophical Studies}, 175\penalty0 (6):\penalty0 1441--1464,
  June 2018.
\newblock ISSN 0031-8116, 1573-0883.
\newblock \doi{10.1007/s11098-017-0919-0}.

\bibitem[Larivière et~al.(2013)Larivière, Ni, Gingras, Cronin, and
  Sugimoto]{lariviere_bibliometrics:_2013}
Vincent Larivière, Chaoqun Ni, Yves Gingras, Blaise Cronin, and Cassidy~R.
  Sugimoto.
\newblock Bibliometrics: {Global} gender disparities in science.
\newblock \emph{Nature}, 504\penalty0 (7479):\penalty0 211--213, December 2013.
\newblock ISSN 0028-0836, 1476-4687.
\newblock \doi{10.1038/504211a}.

\bibitem[Huang et~al.(2020)Huang, Gates, Sinatra, and
  Barabasi]{huang_historical_2020}
Junming Huang, Alexander~J. Gates, Roberta Sinatra, and Albert-Lazlo Barabasi.
\newblock Historical comparison of gender inequality in scientific careers
  across countries and disciplines.
\newblock \emph{Proceedings of the National Academy of Sciences}, 117\penalty0
  (9):\penalty0 4609--4616, March 2020.
\newblock ISSN 0027-8424, 1091-6490.
\newblock \doi{10.1073/pnas.1914221117}.

\bibitem[Ferber and Brun(2011)]{ferber_gender_2011}
Marianne~A. Ferber and Michael Brun.
\newblock The {Gender} {Gap} in {Citations}: {Does} {It} {Persist}?
\newblock \emph{Feminist Economics}, 17\penalty0 (1):\penalty0 151--158,
  January 2011.
\newblock ISSN 1354-5701, 1466-4372.
\newblock \doi{10.1080/13545701.2010.541857}.

\bibitem[Maliniak et~al.(2013)Maliniak, Powers, and
  Walter]{maliniak_gender_2013}
Daniel Maliniak, Ryan Powers, and Barbara~F. Walter.
\newblock The {Gender} {Citation} {Gap} in {International} {Relations}.
\newblock \emph{International Organization}, 67\penalty0 (4):\penalty0
  889--922, October 2013.
\newblock ISSN 0020-8183, 1531-5088.
\newblock \doi{10.1017/S0020818313000209}.

\bibitem[Caplar et~al.(2017)Caplar, Tacchella, and
  Birrer]{caplar_quantitative_2017}
Neven Caplar, Sandro Tacchella, and Simon Birrer.
\newblock Quantitative evaluation of gender bias in astronomical publications
  from citation counts.
\newblock \emph{Nature Astronomy}, 1\penalty0 (6):\penalty0 0141, June 2017.
\newblock ISSN 2397-3366.
\newblock \doi{10.1038/s41550-017-0141}.

\bibitem[Fang et~al.(2000)Fang, Moy, Colburn, and Hurley]{fang_racial_2000}
Di~Fang, Ernest Moy, Lois Colburn, and Jeanne Hurley.
\newblock Racial and {Ethnic} {Disparities} in {Faculty} {Promotion} in
  {Academic} {Medicine}.
\newblock \emph{JAMA}, 284\penalty0 (9):\penalty0 1085--1092, September 2000.
\newblock ISSN 0098-7484.
\newblock \doi{10.1001/jama.284.9.1085}.

\bibitem[Petersen et~al.(2014)Petersen, Fortunato, Pan, Kaski, Penner, Rungi,
  Riccaboni, Stanley, and Pammolli]{petersen_reputation_2014}
Alexander~M. Petersen, Santo Fortunato, Raj~K. Pan, Kimmo Kaski, Orion Penner,
  Armando Rungi, Massimo Riccaboni, H.~Eugene Stanley, and Fabio Pammolli.
\newblock Reputation and impact in academic careers.
\newblock \emph{Proceedings of the National Academy of Sciences}, 111\penalty0
  (43):\penalty0 15316--15321, October 2014.
\newblock ISSN 0027-8424, 1091-6490.
\newblock \doi{10.1073/pnas.1323111111}.

\bibitem[Way et~al.(2019)Way, Morgan, Larremore, and
  Clauset]{way_productivity_2019}
Samuel~F. Way, Allison~C. Morgan, Daniel~B. Larremore, and Aaron Clauset.
\newblock Productivity, prominence, and the effects of academic environment.
\newblock \emph{Proceedings of the National Academy of Sciences}, 116\penalty0
  (22):\penalty0 10729--10733, May 2019.
\newblock ISSN 0027-8424, 1091-6490.
\newblock \doi{10.1073/pnas.1817431116}.

\bibitem[Joels and Mason(2014)]{joels_tale_2014}
Marian Joels and Carol Mason.
\newblock A {Tale} of {Two} {Sexes}.
\newblock \emph{Neuron}, 82\penalty0 (6):\penalty0 1196--1199, June 2014.
\newblock ISSN 08966273.
\newblock \doi{10.1016/j.neuron.2014.05.021}.

\bibitem[noa(2018)]{noauthor_promoting_2018}
Promoting diversity in neuroscience.
\newblock \emph{Nature Neuroscience}, 21\penalty0 (1):\penalty0 1--1, January
  2018.
\newblock ISSN 1097-6256, 1546-1726.
\newblock \doi{10.1038/s41593-017-0052-6}.

\bibitem[Schrouff et~al.(2019)Schrouff, Pischedda, Genon, Fryns, Pinho,
  Vassena, Liuzzi, and Ferreira]{schrouff_gender_2019}
Jessica Schrouff, Doris Pischedda, Sarah Genon, Gregory Fryns, Ana~Luísa
  Pinho, Eliana Vassena, Antonietta~G. Liuzzi, and Fabio~S. Ferreira.
\newblock Gender bias in (neuro)science: {Facts}, consequences, and solutions.
\newblock \emph{European Journal of Neuroscience}, March 2019.
\newblock ISSN 0953816X.
\newblock \doi{10.1111/ejn.14397}.

\bibitem[Chakravartty et~al.(2018)Chakravartty, Kuo, Grubbs, and
  McIlwain]{chakravartty_communicationsowhite_2018}
Paula Chakravartty, Rachel Kuo, Victoria Grubbs, and Charlton McIlwain.
\newblock \#{CommunicationSoWhite}.
\newblock \emph{Journal of Communication}, 68\penalty0 (2):\penalty0 254--266,
  April 2018.
\newblock ISSN 0021-9916, 1460-2466.
\newblock \doi{10.1093/joc/jqy003}.

\bibitem[Thiem et~al.(2018)Thiem, Sealey, Ferrer, Trott, and
  Kennison]{thiem_just_2018}
Yannik Thiem, Kris~F. Sealey, Amy~E. Ferrer, Adriel~M. Trott, and Rebecca
  Kennison.
\newblock Just {Ideas}? {The} {Status} and {Future} of {Publication} {Ethics}
  in {Philosophy}: {A} {White} {Paper}.
\newblock Technical report, 2018.

\bibitem[Dion et~al.(2018)Dion, Sumner, and Mitchell]{dion_gendered_2018}
Michelle~L. Dion, Jane~Lawrence Sumner, and Sara~McLaughlin Mitchell.
\newblock Gendered {Citation} {Patterns} across {Political} {Science} and
  {Social} {Science} {Methodology} {Fields}.
\newblock \emph{Political Analysis}, 26\penalty0 (3):\penalty0 312--327, July
  2018.
\newblock ISSN 1047-1987, 1476-4989.
\newblock \doi{10.1017/pan.2018.12}.

\bibitem[Rossiter(1993)]{rossiter_matthew_1993}
Margaret~W. Rossiter.
\newblock The {Matthew} {Matilda} {Effect} in {Science}.
\newblock \emph{Social Studies of Science}, 23\penalty0 (2):\penalty0 325--341,
  May 1993.
\newblock ISSN 0306-3127, 1460-3659.
\newblock \doi{10.1177/030631293023002004}.

\bibitem[Selisker(2015)]{selisker_bechdel_2015}
Scott Selisker.
\newblock The {Bechdel} {Test} and the {Social} {Form} of {Character}
  {Networks}.
\newblock \emph{New Literary History}, 46\penalty0 (3):\penalty0 505--523,
  2015.
\newblock ISSN 1080-661X.
\newblock \doi{10.1353/nlh.2015.0024}.

\bibitem[Garcia et~al.(2014)Garcia, Weber, and
  Garimella]{garcia_asymmetries2014}
David Garcia, Ingmar Weber, and Venkata Garimella.
\newblock Gender asymmetries in reality and fiction: The bechdel test of social
  media.
\newblock \emph{Proceedings of the 8th International Conference on Weblogs and
  Social Media, ICWSM 2014}, 04 2014.

\bibitem[Mitchell et~al.(2013)Mitchell, Lange, and
  Brus]{mitchell_gendered_2013}
Sara~McLaughlin Mitchell, Samantha Lange, and Holly Brus.
\newblock Gendered {Citation} {Patterns} in {International} {Relations}
  {Journals}1.
\newblock \emph{International Studies Perspectives}, 14\penalty0 (4):\penalty0
  485--492, 2013.
\newblock ISSN 1528-3577.
\newblock \doi{10.1111/insp.12026}.

\bibitem[Bergstrom et~al.(2008)Bergstrom, West, and
  Wiseman]{bergstrom_eigenfactortm_2008}
Carl~T. Bergstrom, Jevin~D. West, and Marc~A. Wiseman.
\newblock The {EigenfactorTM} {Metrics}.
\newblock \emph{Journal of Neuroscience}, 28\penalty0 (45):\penalty0
  11433--11434, November 2008.
\newblock ISSN 0270-6474, 1529-2401.
\newblock \doi{10.1523/JNEUROSCI.0003-08.2008}.

\bibitem[Blevins and Mullen(2015)]{blevins_jane_2015}
Cameron Blevins and Lincoln Mullen.
\newblock Jane, {John} ... {Leslie}? {A} {Historical} {Method} for
  {Algorithmic} {Gender} {Prediction}.
\newblock \emph{Digital Humanities Quarterly}, 9\penalty0 (3), 2015.

\bibitem[Fausto-Sterling(2000)]{fausto-sterling_sexing_2000}
Anne Fausto-Sterling.
\newblock \emph{Sexing the body: gender politics and the construction of
  sexuality}.
\newblock Basic Books, New York, NY, 1st ed edition, 2000.
\newblock ISBN 978-0-465-07713-7 978-0-465-07714-4.
\newblock OCLC: ocm43468482.

\bibitem[Feder(2014)]{feder_making_2014}
Ellen~K. Feder.
\newblock \emph{Making sense of intersex: changing ethical perspectives in
  biomedicine}.
\newblock Indiana University Press, Bloomington, 2014.
\newblock ISBN 978-0-253-01224-1 978-0-253-01228-9.

\bibitem[Stryker(2008)]{stryker_transgender_2008}
Susan Stryker.
\newblock \emph{Transgender history}.
\newblock Seal studies. Seal Press : Distributed by Publishers Group West,
  Berkeley, CA, 2008.
\newblock ISBN 978-1-58005-224-5.
\newblock OCLC: ocn183914566.

\bibitem[Greenwald et~al.(1998)Greenwald, McGhee, and
  Schwartz]{greenwald_measuring_1998}
Anthony~G. Greenwald, Debbie~E. McGhee, and Jordan L.~K. Schwartz.
\newblock Measuring individual differences in implicit cognition: {The}
  implicit association test.
\newblock \emph{Journal of Personality and Social Psychology}, 74\penalty0
  (6):\penalty0 1464--1480, 1998.
\newblock ISSN 1939-1315, 0022-3514.
\newblock \doi{10.1037/0022-3514.74.6.1464}.

\bibitem[Brewer(1999)]{brewer_psychology_1999}
Marilynn~B. Brewer.
\newblock The {Psychology} of {Prejudice}: {Ingroup} {Love} and {Outgroup}
  {Hate}?
\newblock \emph{Journal of Social Issues}, 55\penalty0 (3):\penalty0 429--444,
  January 1999.
\newblock ISSN 0022-4537, 1540-4560.
\newblock \doi{10.1111/0022-4537.00126}.

\bibitem[Brownstein(2019)]{brownstein_implicit_2019}
Michael Brownstein.
\newblock Implicit bias.
\newblock In \emph{The {Stanford} {Encyclopedia} of {Philsophy}}. E. Zalta
  (Ed.), fall 2019 edition edition, 2019.

\bibitem[Abramo et~al.(2009)Abramo, D'Angelo, and
  Caprasecca]{abramo_contribution_2009}
Giovanni Abramo, Ciriaco~Andrea D'Angelo, and Alessandro Caprasecca.
\newblock The contribution of star scientists to overall sex differences in
  research productivity.
\newblock \emph{Scientometrics}, 81\penalty0 (1):\penalty0 137--156, October
  2009.
\newblock ISSN 0138-9130, 1588-2861.
\newblock \doi{10.1007/s11192-008-2131-7}.

\bibitem[Holman and Morandin(2019)]{holman_researchers_2019}
Luke Holman and Claire Morandin.
\newblock Researchers collaborate with same-gendered colleagues more often than
  expected across the life sciences.
\newblock \emph{PLOS ONE}, 14\penalty0 (4):\penalty0 e0216128, April 2019.
\newblock ISSN 1932-6203.
\newblock \doi{10.1371/journal.pone.0216128}.

\bibitem[Lee et~al.(2019)Lee, Karimi, Wagner, Jo, Strohmaier, and
  Galesic]{lee_homophily_2019}
Eun Lee, Fariba Karimi, Claudia Wagner, Hang-Hyun Jo, Markus Strohmaier, and
  Mirta Galesic.
\newblock Homophily and minority-group size explain perception biases in social
  networks.
\newblock \emph{Nature Human Behaviour}, August 2019.
\newblock ISSN 2397-3374.
\newblock \doi{10.1038/s41562-019-0677-4}.

\bibitem[Aksnes et~al.(2019)Aksnes, Langfeldt, and
  Wouters]{aksnes_citations_2019}
Dag~W. Aksnes, Liv Langfeldt, and Paul Wouters.
\newblock Citations, {Citation} {Indicators}, and {Research} {Quality}: {An}
  {Overview} of {Basic} {Concepts} and {Theories}.
\newblock \emph{SAGE Open}, 9\penalty0 (1):\penalty0 215824401982957, January
  2019.
\newblock ISSN 2158-2440, 2158-2440.
\newblock \doi{10.1177/2158244019829575}.

\bibitem[Henry(2010)]{henry_institutional}
PJ~Henry.
\newblock \emph{Henry, P. J. (2010). Institutional bias. In J. F. Dovidio, M.
  Hewstone, P. Glick, V. M. Esses (Eds.), Handbook of prejudice, stereotyping,
  and discrimination (426-440). London: Sage.}, pages 426--440.
\newblock 01 2010.

\bibitem[Clarke()]{clarke_explicit_nodate}
Jessica~A. Clarke.
\newblock Explicit {Bias}.
\newblock \emph{Northwestern University Law Review}, 113\penalty0 (3):\penalty0
  505--586.

\bibitem[Greenwald and Banaji(1995)]{greenwald_implicit_1995}
Anthony~G. Greenwald and Mahzarin~R. Banaji.
\newblock Implicit social cognition: {Attitudes}, self-esteem, and stereotypes.
\newblock \emph{Psychological Review}, 102\penalty0 (1):\penalty0 4--27, 1995.
\newblock ISSN 1939-1471, 0033-295X.
\newblock \doi{10.1037/0033-295X.102.1.4}.

\bibitem[Conaway and Bethune(2015)]{conaway_implicit_2015}
Wendy Conaway and Sonja Bethune.
\newblock Implicit {Bias} and {First} {Name} {Stereotypes}: {What} are the
  {Implications} for {Online} {Instruction}?
\newblock \emph{Online Learning}, 19\penalty0 (3), March 2015.
\newblock ISSN 2472-5730, 2472-5749.
\newblock \doi{10.24059/olj.v19i3.452}.

\bibitem[Bertrand and Mullainathan(2004)]{bertrand_are_2004}
Marianne Bertrand and Sendhil Mullainathan.
\newblock Are {Emily} and {Greg} {More} {Employable} {Than} {Lakisha} and
  {Jamal}? {A} {Field} {Experiment} on {Labor} {Market} {Discrimination}.
\newblock \emph{American Economic Review}, 94\penalty0 (4):\penalty0 991--1013,
  August 2004.
\newblock ISSN 0002-8282.
\newblock \doi{10.1257/0002828042002561}.

\bibitem[Paludi and Strayer(1985)]{paludi_whats_1985}
Michele~A. Paludi and Lisa~A. Strayer.
\newblock What's in an author's name? {Differential} evaluations of performance
  as a function of author's name.
\newblock \emph{Sex Roles}, 12\penalty0 (3-4):\penalty0 353--361, February
  1985.
\newblock ISSN 0360-0025, 1573-2762.
\newblock \doi{10.1007/BF00287601}.

\bibitem[Jadidi et~al.(2018)Jadidi, Karimi, Lietz, and
  Wagner]{jadidi_gender_2018}
Mohsen Jadidi, Fariba Karimi, Haiko Lietz, and Claudia Wagner.
\newblock {Gender} {Disparities} {in} {Science}? {Dropout}, {Productivity},
  {Collaborations} {and} {Success} {of} {Male} {and} {Female} {Computer}
  {Scientists}.
\newblock \emph{Advances in Complex Systems}, 21\penalty0 (03n04):\penalty0
  1750011, May 2018.
\newblock ISSN 0219-5259, 1793-6802.
\newblock \doi{10.1142/S0219525917500114}.

\bibitem[Posselt(2016)]{posselt_inside_2016}
Julie~R. Posselt.
\newblock \emph{Inside {Graduate} {Admissions}}.
\newblock Harvard University Press, 2016.
\newblock ISBN 978-0-674-08869-6.

\bibitem[Colgan(2017)]{colgan_gender_2017}
Jeff Colgan.
\newblock Gender {Bias} in {International} {Relations} {Graduate} {Education}?
  {New} {Evidence} from {Syllabi}.
\newblock \emph{PS: Political Science \& Politics}, 50\penalty0 (02):\penalty0
  456--460, April 2017.
\newblock ISSN 1049-0965, 1537-5935.
\newblock \doi{10.1017/S1049096516002997}.

\bibitem[Penders(2018)]{penders_ten_2018}
Bart Penders.
\newblock Ten simple rules for responsible referencing.
\newblock \emph{PLOS Computational Biology}, 14\penalty0 (4):\penalty0
  e1006036, April 2018.
\newblock ISSN 1553-7358.
\newblock \doi{10.1371/journal.pcbi.1006036}.

\bibitem[Sumner(2018)]{sumner_gender_2018}
Jane~Lawrence Sumner.
\newblock The {Gender} {Balance} {Assessment} {Tool} ({GBAT}): {A}
  {Web}-{Based} {Tool} for {Estimating} {Gender} {Balance} in {Syllabi} and
  {Bibliographies}.
\newblock \emph{PS: Political Science \& Politics}, 51\penalty0 (02):\penalty0
  396--400, April 2018.
\newblock ISSN 1049-0965, 1537-5935.
\newblock \doi{10.1017/S1049096517002074}.

\bibitem[Gutiérrez~y Muhs et~al.(2012)Gutiérrez~y Muhs, Niemann, González,
  and Harris]{gutierrez_y_muhs_presumed_2012}
Gabriella Gutiérrez~y Muhs, Yolanda~F. Niemann, Carmen~G. González, and
  Angela~P. Harris, editors.
\newblock \emph{Presumed incompetent: the intersections of race and class for
  women in academia}.
\newblock University Press of Colorado, Boulder, Colo, 2012.
\newblock ISBN 978-0-87421-869-5 978-0-87421-870-1.

\bibitem[noa()]{noauthor_up_nodate}
The {UP} {Directory}: {A} directory of philosophers from underrepresented
  groups in philosophy.

\bibitem[Rawls(1999)]{rawls_theory_1999}
John Rawls.
\newblock \emph{A theory of justice}.
\newblock Belknap Press of Harvard University Press, Cambridge, Mass, rev. ed
  edition, 1999.
\newblock ISBN 978-0-674-00077-3 978-0-674-00078-0.

\bibitem[Lamont and Favor(2017)]{lamont_distributive_2017}
Julian Lamont and Christi Favor.
\newblock Distributive {Justice}.
\newblock In \emph{The {Stanford} {Encyclopedia} of {Philosophy}}. Edward N.
  Zalta (Ed.), winter 2017 edition edition, 2017.

\bibitem[Olsaretti(2018)]{olsaretti_idea_2018}
Serena Olsaretti.
\newblock \emph{The {Idea} of {Distributive} {Justice}}, volume~1.
\newblock Oxford University Press, June 2018.
\newblock \doi{10.1093/oxfordhb/9780199645121.013.38}.

\bibitem[Young and Allen(2011)]{young_justice_2011}
Iris~Marion Young and Danielle~S. Allen.
\newblock \emph{Justice and the politics of difference}.
\newblock Princeton University Press, Princeton, N.J, paperback reissue
  edition, 2011.
\newblock ISBN 978-0-691-15262-2.
\newblock OCLC: ocn751237488.

\bibitem[Ahmed(2012)]{ahmed_being_2012}
Sara Ahmed.
\newblock \emph{On being included: racism and diversity in institutional life}.
\newblock Duke University Press, Durham ; London, 2012.
\newblock ISBN 978-0-8223-5221-1 978-0-8223-5236-5.

\bibitem[Walker(2010)]{walker_what_2010}
Margaret~Urban Walker.
\newblock \emph{What is reparative justice?}
\newblock Number 2010 in Aquinas lecture. Marquette University Press,
  Milwaukee, Wis, 2010.
\newblock ISBN 978-0-87462-177-8.
\newblock OCLC: ocn503596039.

\bibitem[Anderson(2010)]{anderson_imperative_2010}
Elizabeth Anderson.
\newblock \emph{The imperative of integration}.
\newblock Princeton University Press, Princeton, N.J, 2010.
\newblock ISBN 978-0-691-13981-4.
\newblock OCLC: ocn587248993.

\bibitem[Yang et~al.(2019)Yang, Chawla, and Uzzi]{yang_networks_2019}
Yang Yang, Nitesh~V. Chawla, and Brian Uzzi.
\newblock A network's gender composition and communication pattern predict
  women's leadership success.
\newblock \emph{Proceedings of the National Academy of Sciences}, 116\penalty0
  (6):\penalty0 2033--2038, February 2019.
\newblock ISSN 0027-8424, 1091-6490.
\newblock \doi{10.1073/pnas.1721438116}.

\bibitem[AlShebli et~al.(2018)AlShebli, Rahwan, and
  Woon]{alshebli_preeminence_2018}
Bedoor~K. AlShebli, Talal Rahwan, and Wei~Lee Woon.
\newblock The preeminence of ethnic diversity in scientific collaboration.
\newblock \emph{Nature Communications}, 9\penalty0 (1):\penalty0 5163, December
  2018.
\newblock ISSN 2041-1723.
\newblock \doi{10.1038/s41467-018-07634-8}.

\bibitem[Uhly et~al.(2015)Uhly, Visser, and Zippel]{uhly_gendered_2015}
K.M. Uhly, L.M. Visser, and K.S. Zippel.
\newblock Gendered patterns in international research collaborations in
  academia.
\newblock \emph{Studies in Higher Education}, pages 1--23, September 2015.
\newblock ISSN 0307-5079, 1470-174X.
\newblock \doi{10.1080/03075079.2015.1072151}.

\bibitem[Zippel(2017)]{zippel_women_2017}
Kathrin~S. Zippel.
\newblock \emph{Women in global science: advancing academic careers through
  international collaboration}.
\newblock Stanford University Press, Stanford, California, 2017.
\newblock ISBN 978-1-5036-0039-3 978-1-5036-0149-9.

\bibitem[Bergstrom(2007)]{bergstrom_eigenfactor:_2007}
Carl~T. Bergstrom.
\newblock Eigenfactor: {Measuring} the value and prestige of scholarly
  journals.
\newblock \emph{College \& Research Libraries News}, 68\penalty0 (5):\penalty0
  314--316, May 2007.
\newblock ISSN 2150-6698, 0099-0086.
\newblock \doi{10.5860/crln.68.5.7804}.

\bibitem[Toth et~al.(2014)Toth, Durham, Kantarcioglu, Xue, and
  Malin]{toth_soempi:_2014}
Csaba Toth, Elizabeth Durham, Murat Kantarcioglu, Yuan Xue, and Bradley Malin.
\newblock {SOEMPI}: {A} {Secure} {Open} {Enterprise} {Master} {Patient} {Index}
  {Software} {Toolkit} for {Private} {Record} {Linkage}.
\newblock \emph{AMIA ... Annual Symposium proceedings. AMIA Symposium},
  2014:\penalty0 1105--1114, 2014.
\newblock ISSN 1942-597X.

\bibitem[King et~al.(2017)King, Bergstrom, Correll, Jacquet, and
  West]{king_men_2017}
Molly~M. King, Carl~T. Bergstrom, Shelley~J. Correll, Jennifer Jacquet, and
  Jevin~D. West.
\newblock Men {Set} {Their} {Own} {Cites} {High}: {Gender} and {Self}-citation
  across {Fields} and over {Time}.
\newblock \emph{Socius: Sociological Research for a Dynamic World}, 3:\penalty0
  237802311773890, December 2017.
\newblock ISSN 2378-0231, 2378-0231.
\newblock \doi{10.1177/2378023117738903}.

\bibitem[Wood(2017)]{mgcv}
Simon~N. Wood.
\newblock \emph{Generalized Additive Models: An Introduction with R}.
\newblock Chapman and Hall/CRC, 2 edition, 2017.

\bibitem[Holm(1979)]{holm_simple_1979}
Sture Holm.
\newblock A simple sequentially rejective multiple test procedure.
\newblock \emph{Scandinavian Journal of Statistics}, 6\penalty0 (2):\penalty0
  65--70, 1979.

\end{thebibliography}
\end{document}